\documentclass[ twocolumn,amsmath,amssymb,aps,floats,floatfix,nofootinbib]{revtex4}
\usepackage[colorlinks, linkcolor=blue, anchorcolor=red, citecolor=green, colorlinks=true]{hyperref}
\usepackage{amssymb}
\usepackage{amsmath}
\usepackage{graphicx}
\usepackage{indentfirst}
\usepackage{subfigure}
\usepackage{ulem}
\usepackage{bm}
\usepackage{xspace}
\usepackage{epstopdf}
\makeatother
\allowdisplaybreaks 

\newcommand{\pks}{PKS\,2155$-$304\xspace}
\newcommand{\pg}{PG\,1553$+$113\xspace}
\newcommand{\hess}{H.E.S.S.\xspace}
\newcommand{\phasetwo}{\hess~II\xspace}

\begin{document}
\title{The implications of the axion like particle from the Fermi-LAT and \hess observations of \pg and \pks}
\author{Jun-Guang Guo$^{1,2}$}
\author{Hai-Jun Li$^{1,2}$}
\author{Xiao-Jun Bi$^{1,2}$}
\author{Su-Jie Lin$^{1}$}
\author{Peng-Fei Yin$^{1}$}
\affiliation{$^{1}$Key Laboratory of Particle Astrophysics, Institute of High Energy Physics, Chinese Academy of Sciences, Beijing 100049, China   \\
$^{2}$School of Physics, University of Chinese Academy of Sciences, Beijing 100049, China}

\begin{abstract}
We investigate the axion like particle (ALP)-photon oscillation effect in the high energy $\gamma$-ray spectra of \pg and \pks measured by Fermi-LAT and \hess.
The choice of extragalactic background light (EBL) model, which induces the attenuate effect in observed $\gamma$-ray spectra, would affect the ALP implication.
For the ordinary EBL model that prefers a null hypothesis, we set constraint on the ALP-photon coupling constant at 95\% C.L. as $g_{a\gamma}\lesssim 5\times 10^{-11} ~\rm{GeV}^{-1}$ for the ALP mass $\sim 10$ neV.
We also consider the CIBER observation of the cosmic infrared radiation, which shows an excess at the wave wavelength of $\sim 1~\mu$m after the substraction of foregrounds. The high energy gamma-rays from extragalactic sources at high redshifts would suffer from a more significant attenuate effect caused by this excess.
In this case, we find that the ALP-photon oscillation would improve the fit to the observed spectra of \pks and \pg and find a favored parameter region at 95\% C.L..
\end{abstract}
\maketitle

\section{Introduction}

Axion is a light pseudo-Goldstone boson proposed to solve the strong CP problem in QCD
\cite{Peccei:1977hh, Weinberg:1977ma, Peccei:2006as}.
Many new physics models beyond the standard model also suggest the existence of axion-like particles (ALPs) \cite{Cicoli:2012sz, DiLella:2000dn}. These particles may play an important role in the evolution of the universe and have rich phenomenology in high energy and astrophysics experiments.
Considering the effective coupling between the ALP and photons, many investigations for the ALP-photon conversion effect have been performed \cite{sikivie1983experimental, asztalos2010squid, carosi2013probing, Graham:2015ouw, Majumdar:2017vcx, Galanti:2018myb}. For instance, the CAST experiment investigated the photon signal induced by the ALPs from the Sun and has set a stringent constraint on the ALP-photon coupling as $g_{a\gamma}\leq 6.6\times 10^{-11}~{\rm GeV^{-1}}$ \cite{Anastassopoulos:2017ftl}.

It is promising to explore ALP through the ALP-photon oscillation effect in high energy astrophysical processes \cite{Raffelt:1990yz}.
The initial high energy photons emitted from the astrophysical source may be converted into ALP by the external magnetic field around the source \cite{Raffelt:1987im,Mirizzi:2009aj,DeAngelis:2011id}. Then ALPs propagate in the extragalactic space without energy loss, compared with high energy photons which could interact with the extragalactic background light (EBL) background. Finally, these ALPs can be converted into detectable high energy photons by the Galactic magnetic field. Therefore, it is expected that the ALP-photon oscillation would reduce the attenuation effect of high energy photons from distant sources and affect the final photon spectra.

Using the data from the observations of high energy photons, many studies on the ALP-photon conversion have been performed in the literature \cite{Belikov:2010ma, Abramowski:2013oea, Reesman:2014ova, Payez:2014xsa, Berenji:2016jji, TheFermi-LAT:2016zue, Meyer:2016wrm, Liang:2018mqm, Zhang:2018wpc, Libanov:2019fzq, Long:2019nrz, Majumdar:2017vcx, Galanti:2018myb, Troitsky:2015nxa, Csaki:2003ef, DeAngelis:2007dqd, DeAngelis:2011id, Simet:2007sa, Fairbairn:2009zi, Meyer:2013pny, Dominguez:2011xy,
Mirizzi:2009aj, Mirizzi:2007hr, de2009photon, SanchezConde:2009wu, Kohri:2017ljt, Bi:2020ths}. 
For instance, the $\gamma$-ray spectra from the sources NGC 1275 and \pks at high redshifts measured by Fermi-LAT have been used to set constraints on $g_{a\gamma}$ in Ref.~\cite{TheFermi-LAT:2016zue} and Ref.~\cite{Zhang:2018wpc}, respectively. Compared with the detectable energy range of Fermi-LAT $\sim 0.1$ GeV-300 GeV, imaging atmospheric Cherenkov telescopes detect high energy $\gamma$-ray above $\mathcal{O}(10^2)$ GeV, which would open a different window for the ALP research. For instance, The data of \pks from \hess I observation has been used to search for ALP in Ref.~\cite{Abramowski:2013oea}.

Compared with \hess I, \phasetwo with the fifth telescope added in 2012 is sensitive to the $\gamma-$ray spectra at lower energies. The \phasetwo measurements of the very high energy (VHE, E$\gtrsim 100{\rm GeV}$) gamma-ray spectra of two extragalactic sources \pks and \pg have been reported in Ref.~\cite{Aharonian:2016ria}. These two sources are high frequency peaked BL Lac objects with high statistics in the VHE $\gamma$-ray sky. Since they are located at high red-shifts (z=0.116 and 0.49 for \pks and \pg, respectively), the EBL attenuation effects for their spectra are expected to be significant. Consequently, the measurements of VHE spectra are suitable to detect the ALP-photon oscillation, which can compensate the EBL attenuation effect.

This paper is organized as follows. In Section.~\ref{section_EBL}, we describe the EBL attenuation effect and introduce the two EBL models adopted in this work.
In Section.~\ref{section_apo}, we introduce the ALP-photon oscillation effect in the extragalactic source and Milky Way.
In Section.~\ref{section_method}, we describe our fitting method to the observed $\gamma$-ray spectra.
In Section.~\ref{section_results}, we investigate the implication of ALP in the data for the two EBL models.
The conclusions are given in Section.~\ref{section_Con}.

\section{EBL attenuation effect}
\label{section_EBL}

Before entering the Galaxy, high energy $\gamma$-ray would interact with the EBL and loss energy through the pair production $\gamma + \gamma_{\rm EBL} \to e^+ + e^-$. This attenuation effect can be described by the factor of $e^{-\tau\left(E_\gamma, z_0\right)}$, where $\tau\left(E_\gamma, z_0\right)$ is the optical length for the source at redshift of $z_0$
\begin{align*}
    \tau\left(E_\gamma,z_0\right)=&\int_0^{z_0}\frac{{\rm d}z}{\left(1+z\right)H\left(z\right)} \\
                                  &\int_{E_{\gamma}\geq E_{\rm th}}{\rm d}\omega\frac{{\rm d}n}{{\rm d}\omega}\bar{\sigma}\left(E_\gamma,\omega,z\right),
\label{eq:ebl_dis}
\end{align*}
where $E_{\rm th}$ is the threshold energy of the pair production, $\bar{\sigma}$ is the integral cross section of the pair production, and
${\rm d}n/{\rm d}\omega$ is the proper number density per unit energy of the EBL at redshift of $z$.

The main contributions to the EBL at wavelengths from UV to IR are expected to be the starlight and dust re-radiation, accumulated over the history of the universe. In order to predict the EBL, the detailed modeling of the evolution of galaxy populations is needed.
Many EBL models based on empirical or semi-analytical approaches have been developed in the literature ~\cite{Franceschini:2008tp,Finke:2009xi,Dominguez:2010bv,Gilmore:2011ks,Stecker:2016fsg}. According to the method dealing with the evolution of the galaxy populations and the EBL, these models can be classified into four types \cite{Dominguez:2010bv}. Using the cosmological survey data from a variety of ground-based experiments and space telescopes, Franceschini et al. built a backward evolutionary model to extrapolate the evolution of the EBL~\cite{Franceschini:2008tp} (hereafter FRV08 model). The observed galaxy luminosity functions are used to derive the contributions from different galaxy populations based on morphology. In our work, we adopt this EBL model to compute the gamma-ray attenuation effect.

\begin{figure}[htb]
\centering
\includegraphics[width=0.5\textwidth]{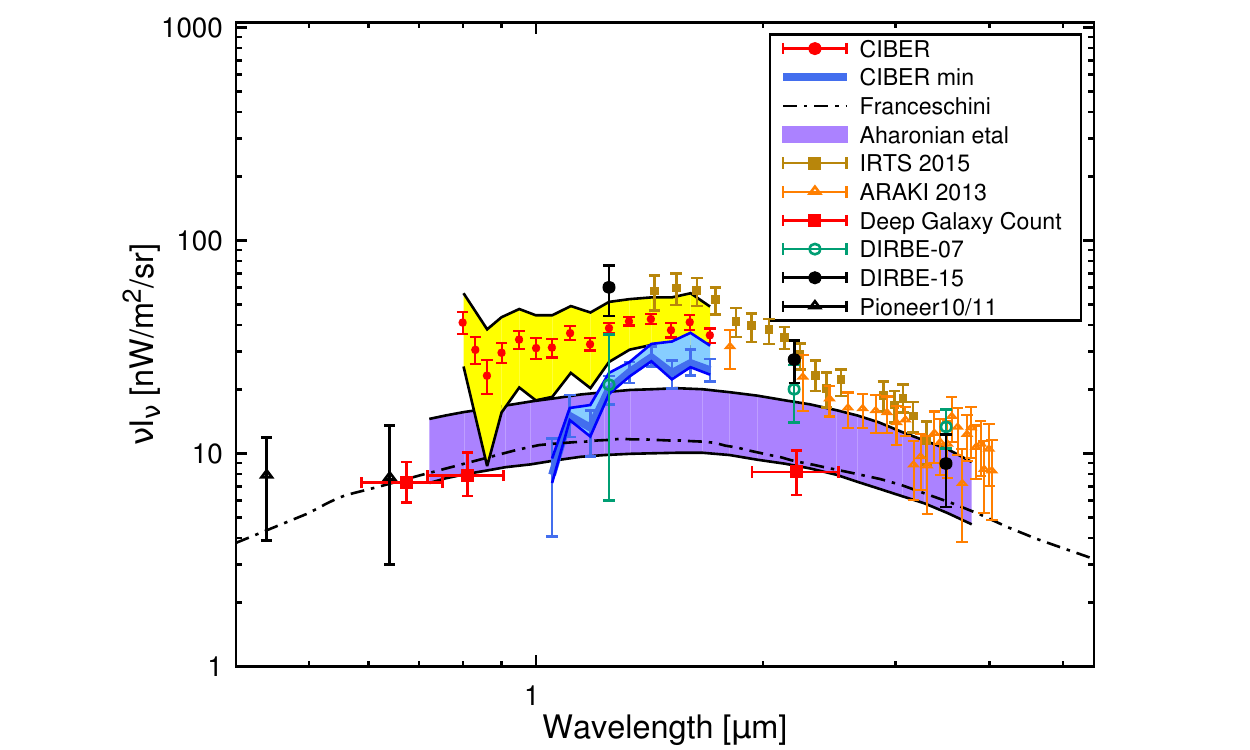}
\caption{EBL spectra from the CIBER~\cite{Matsuura:2017lub}, IRTS~\cite{Matsumoto:2015fma}, AKARI~\cite{Tsumura:2013iza}, COBE/DIRBE \cite{Levenson:2007is,Sano:2015bsa}, and Pioneer 10/11 \cite{Matsuoka:2011hb} results.
Also shown are the FRV08 EBL model (dashed dotted line) provided in Ref.~\cite{Franceschini:2008tp}.}
\label{fig:cib}
\end{figure}

It is difficult to directly measure the EBL spectrum due to the bright foregrounds, such as the zodiacal light which is sunlight scattered by the interplanetary dust. Some efforts have been made to directly derive the EBL at near-IR wavelengths by subtracting the foregrounds from the data \cite{Hauser:1998ri,Matsumoto:2004dx,Tsumura:2013iza,Matsumoto:2015fma,Sano:2015bsa,Matsuura:2017lub}. It is interesting to note that many analyses suggest an isotropic excess in the range of
$\sim 1-4~ \mu \rm{m}$ compared with the integrated light from galaxies predicted from deep galaxy counts and theoretical models.
Recently, Ref.~\cite{Matsuura:2017lub} reported the derived EBL in the wavelength range of 0.8$-$1.7$\mu$m from the CIBER observation, which is shown in Fig.~\ref{fig:cib}.
The absolute brightness of the derived EBL is highly dependent on the subtraction of zodiacal light. Assuming the Kelsall zodiacal light model \cite{Kelsall:1998bq}, the residual brightness is 42.7$^{+11.9}_{-10.6} \rm{nW}\cdot \rm{m}^{-2}\rm{sr}^{-1}$ at $1.4 ~\mu\rm{m}$. Using a model independent method for the substraction, the derived minimum EBL brightness is 28.7$^{+5.1}_{-3.3} \rm{nW}\cdot \rm{m}^{-2}\rm{sr}^{-1}$ at $1.4 ~\mu\rm{m}$, which still exceeds the theoretical results.

This excess may be explained by a new foreground component or a new EBL component. For instance, the radiation from the Population III stars at redshifts $\sim 10-20$ may contribute to this component \cite{Kashlinsky:2004xx}. Some studies also investigate the possibility that this component is produced by the decay of ALP \cite{Kohri:2017oqn, Kalashev:2018bra, Korochkin:2019qpe}.

It is expected that the high energy $\gamma$-ray spectrum of the astrophysical source at high redshifts would suffer from a significant attenuation effect after considering the excess. Therefore, the VHE gamma-ray observations set constraints on the EBL. There is a conflict between the results from these analyses and the directly derived EBL at $\mathcal{O}(1) \mu\rm{m}$ (see e.g. Ref.~\cite{Acciari:2019zgl,Abeysekara:2019ybp}). Ref.~\cite{Kohri:2017ljt} found that this conflict can be reconciled by the oscillation between the photons and ALP and find a ALP parameter region favored by observations. In this work, we consider the EBL model with an excess at $\mathcal{O}(1)$$~\mu$m based on the CIBER result \cite{Matsuura:2017lub} (hereafter Ciber model) and investigate the ALP implication using a different method to calculate of ALP-photon conversion compared with Ref.~\cite{Kohri:2017ljt}. We incorporate the CIBER result into the FRV08 spectrum at present and only consider the redshift evolution for this excess at $z\sim 0-0.5$.

\section{ALP-photon oscillation in propagation}
\label{section_apo}

In this section we describe the ALP-photon oscillation effect in propagation.
The ALP-photon conversation arises from the effective coupling between the ALP and photons through the triangle graph with internal fermion lines.
The effective lagrangian is written as
\begin{align}
    \mathcal{L}_{a\gamma}&=-\frac{1}{4}g_{a\gamma}a F_{\mu\nu}\tilde{F}^{\mu\nu}
    =g_{a\gamma}a \textbf{E}\cdot\textbf{B},
\end{align}
where $a$ is the ALP field, $F_{\mu\nu}$ is the electromagnetic field tensor, $\tilde{F}^{\mu\nu}$ is the dual tensor,
and \textbf{E} and \textbf{B} represent the electric and magnetic field, respectively.
The ALP-photon beam can be described by $\Psi = (A_1, A_2, a)^T$, where
$A_1$ and $A_2$ represent the photon transverse polarization states along two orthogonal directions $\hat{\textbf{x}}_1$ and $\hat{\textbf{x}}_2$, respectively.
The ALP-photon beam obeys the Von-Neumann-like equation \cite{Mirizzi:2009aj,DeAngelis:2011id}
\begin{eqnarray}
\frac{{\rm d}\rho}{{\rm d}s}=\left[\rho,\mathcal{M}_0\right],
\end{eqnarray}
where $s$ represents the traveling distance of the ALP-photon beam along the propagation direction $\hat{\textbf{x}}_3 \equiv \hat{\textbf{x}}_1 \times \hat{\textbf{x}}_2$,
$\mathcal{M}_0$ is the mixing matrix, and $\rho$ is the density matrix of the beam $\rho=\Psi \otimes \Psi^\dagger$.
$\mathcal{M}_0$ is only related with the transverse magnetic field $\textbf{\rm {B}}_\perp$.

Assuming that $\textbf{\rm {B}}_\perp$ is aligned along $\hat{\textbf{x}}_2$,
the mixing matrix is\cite{Raffelt:1987im, Mirizzi:2007hr}
\begin{eqnarray}
\mathcal{M}_0={
	\left[\begin{array}{ccc}
    \Delta_{\rm {pl}} & 0 & 0\\
    0 & \Delta_{\rm {pl}} & \Delta_{a\gamma}\\
	0 & \Delta_{a\gamma} & \Delta_{aa}
	\end{array}
	\right]},
\end{eqnarray}
where $\Delta_{\rm{pl}}=-\omega^2_{\rm {pl}}/(2E)$ represents the plasma effect with the plasma frequency $\omega_{\rm{pl}}$ and photon energy $E$,
$\Delta_{aa} = - m^2_a / (2E)$ represents the kinetic term for the ALP with mass of $m_a$, and
$\Delta_{a\gamma}$ is the ALP-photon coupling term $ g_{a\gamma}B_\perp/2$. The Faraday rotation and QED vacuum polarization effect are neglected here.

If $B_\perp$ is not aligned along $\hat{\textbf{x}}_2$, the mixing matrix becomes
\begin{eqnarray}
\label{eq:m_sol}
\mathcal{M}=V\left(\psi\right)\mathcal{M}_0V^\dagger\left(\psi\right),
\end{eqnarray}
with
\begin{eqnarray}
V\left(\psi\right)=
\left[
\begin{array}{ccc}
{\rm cos}\psi & {\rm sin}\psi & 0 \\
-{\rm sin}\psi & {\rm cos}\psi & 0 \\
0 & 0 & 1
\end{array}
\right],
\end{eqnarray}
where $\psi$ is the angle between $B_\perp$ and $\hat{\textbf{x}}_2$. In the general case, the magnetic field of the astrophysical system changes its direction along the propagation direction $\hat{\textbf{x}}_3 $.
In order to describe this effect, the propagation path is divided into $n$ small regions. In each region, the magnetic field can be approximately treated as a constant.
The transfer matrix $T\left(s\right)$ is given by
\begin{eqnarray}
    T\left(s\right)=\prod^{n}_{i}\mathcal{M}(i)
\end{eqnarray}
where $\mathcal{M}(i)$ represents the mixing matrix in the i-th region.

In this work, we consider the high energy $\gamma$-ray spectra from two extragalactic sources \pks and \pg, which are high-frequency peaked BL Lac objects. It is known that BL Lac objects are hosted in elliptical galaxies. However, it is not easy to determine the exact cluster environments around these objects. There are evidences that some BL Lac objects are harboured in small galaxy groups or clusters~\cite{pesce1995environmental, Falomo:2014yya}.
Some studies~\cite{falomo1993environment,Farina:2015rwa} also show that \pks is located at the center of a galaxy cluster.
Thus it can be expected that the high energy photons emitted from the BL Lac objects oscillate with ALP in the inter-cluster magnetic field (ICMF).
The strength of regular magnetic field in the galaxy cluster ranges from $\sim$1~$\mu $G to 10~$\mu $G \cite{Carilli:2001hj}.
We assume that ICMF is a Gaussian turbulent field as Ref.~\cite{Meyer:2014epa}, whose mean value is zero and variance is $\sigma_B$. Since no concrete ICMF model is available, we randomly generate the configuration of ICMF following Ref.~\cite{TheFermi-LAT:2016zue}. 100 realizations of ICMF for each source are taken in the analysis.
The fiducial parameters of ICMF are adopted as Ref.~\cite{Zhang:2018wpc},
where the typical $\sigma_B$ is taken to be 3~$\mu$G.

In this analysis, we do not consider the impact of the magnetic field in the extragalactic space. Some researches show that the upper limit of its strength is $\mathcal{O}(1)$ nG\cite{Pshirkov:2015tua}, but the exact value remains unclear. Thus only the EBL attenuation effect is taken into account for the $\gamma$-ray propagation in the extragalactic space.

The ALP-photon oscillation would also occur in the Milky Way. The galactic magnetic field consists of two components: the random component in small scale and the regular component in large scale. The impact of the random component is neglected here due to the short coherent length. For the regular component, we take the model in Ref.~\cite{Jansson:2012rt}.

The final transfer matrix consists of the contributions from three regions
\begin{eqnarray}
    T\left(s\right) = T_{{\rm MW}}T_{{\rm EBL}}T_{{\rm ICMF}},
\end{eqnarray}
where $T_{\rm MW}$, $T_{\rm EBL}$ and $T_{\rm ICMF}$ are the transfer functions in the
Galactic magnetic field, EBL, and ICMF, respectively.
The density matrix can be solved by
\begin{eqnarray}
\rho\left(s\right)=T\left(s\right)\rho\left(0\right)T^\dagger\left(s\right),
\end{eqnarray}
where $\rho\left(0\right)$ represents the density matrix for the initial beam,
which is assumed to be a pure photon beam without polarization $\rho(0)=\frac{1}{2}~ {\rm diag}(1, 1, 0)$.
The survival probability of photons in the final beam is given by
$P_{\gamma}=\rho_{1}+\rho_{2}$, where $\rho_{1}\left(s\right)$ and $\rho_{2}\left(s\right)$ are the first and second diagonal elements in the density matrix $\rho\left(s\right)$, respectively.

\begin{figure}[htb]
\includegraphics[width=0.50\textwidth]{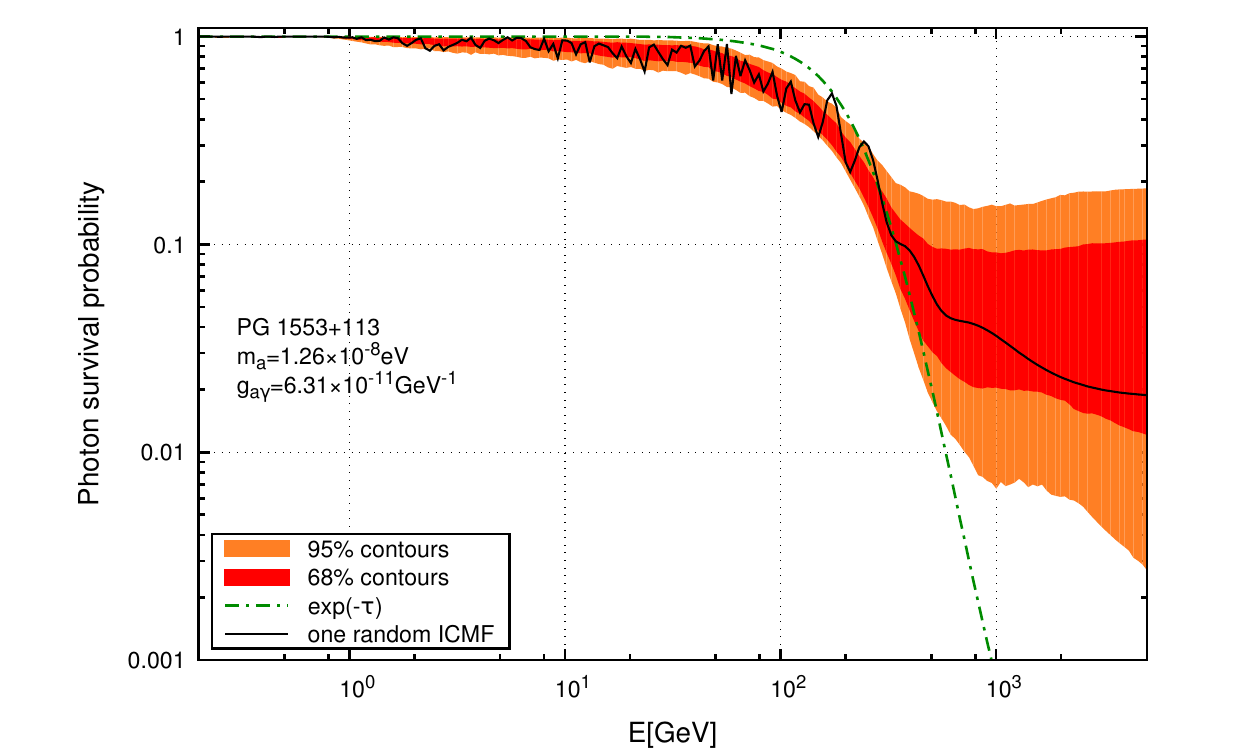}
\caption{Survival probability of $\gamma$-ray emitted from \pg with $m_a=1.26 \times 10^{-8}$ eV and $g_{a\gamma}=6.31\times 10^{-11}$ GeV$^{-1}$ for the FRV08 model.
The solid line represents the result for one randomly selected realization of ICMF.
The red (yellow) band represents the 68\% (95\%) bands for 100 realizations of ICMF.
The dotted dashed line represents the survival probability of $\gamma$-ray without the ALP-photon oscillation.}
\label{fig:sp_band}
\end{figure}

In Fig.~\ref{fig:sp_band}, we show the photon survival probability with $m_a=1.26 \times 10^{-8}$ eV and $g_{a\gamma}=6.31\times 10^{-11}$ GeV$^{-1}$ as a function of the photon energy for one ICMF realization of \pg. The EBL model is taken to be the FRV08 model \cite{Franceschini:2008tp}. In order to describe the impact of the randomness of ICMF, we also plot the 68$\%$ and 95$\%$ bands of the photon survival probability using 100 generated realizations of ICMF. The photon survival probabilities with only the EBL attenuation effect are also shown for comparison. We can see that the oscillation effect becomes significant above $\mathcal{O}(10)$ GeV. For VHE $\gamma$-rays above $\mathcal{O}(300)$ GeV, the oscillation effect induces a larger survival probability in comparison with the pure EBL absorption effect.

\section{Analysis Method}
\label{section_method}

In this work, we assume that the initial $\gamma$-ray spectrum of \pks is described by the broken power Law with a transition region\cite{meyer2013opacity},
\begin{eqnarray}
    F(E)=N(E/E_c)^{-\Gamma _1}(1+(E/E_{{\rm break}})^{f})^{(\Gamma _1-\Gamma _2)/f}.
\end{eqnarray}
The spectrum of \pg is fitted with a logarithmic parabola function,
\begin{eqnarray}
    F(E)=N(E/E_0)^{-\alpha - \beta \cdot \ln (E/E_0)},
\end{eqnarray}
where $N$, $\Gamma _1$, $E_{{\rm break}}$, $f$, $\Gamma _2$, $\alpha$, $\beta$, and $E_0$ are taken to be free parameters and $E_c$ is a normalization parameter. Compared with some other spectral forms, these two spectra can provide a better fit to the data under the null hypothesis. Then we derive the expected $\gamma$-ray spectra by using the photon survival probability and fit the experimental data. The observed spectra given by the \phasetwo (CT5 mono) and Fermi-LAT observations\cite{Aharonian:2016ria} are used in this analysis. In order to include the energy resolution of the experiment, the expected $\gamma$-ray flux in an energy bin between $E_1$ and $E_2$ is smeared as
\begin{eqnarray}
\frac{{\rm d}\Phi}{{\rm d}E}=\frac{\int^{E_2}_{E_1} {\rm d}E\int^{\infty}_{0} S(E^\prime, E)F(E^\prime){\rm d}E^\prime}{E_2-E_1}
\end{eqnarray}
where $E$ and $E^\prime$ are the measured and original photon energies, respectively, and $S(E^\prime, E)$ is the gaussian function with a standard deviation of $\sigma$. Here the energy resolutions of \phasetwo and Fermi-LAT are adopted to be 25\% \cite{Aharonian:2016ria} and 15\% \footnote{\url{https://fermi.gsfc.nasa.gov/ssc/data/analysis/documentation/Cicerone/Cicerone_Introduction/LAT_overview.html}}, respectively.

After integrating the observed energy, the expected photon flux is
\begin{eqnarray}
\frac{{\rm d}\Phi}{{\rm d}E}=\frac{\int^{\infty}_{0} A(E^\prime, E_1, E_2)F(E^\prime)dE^\prime}{E_2-E_1},
\end{eqnarray}
where A($E^\prime$, $E_1$, $E_2$) is given by
\begin{equation}
    A(E^\prime, E_1, E_2)=\frac{1}{2}\left[{\rm erf}\left(\frac{E_2-E^\prime}{\sqrt{2}\sigma}\right)-{\rm erf}\left(\frac{E_1-E^\prime}{\sqrt{2}\sigma}\right)\right],
\end{equation}
where $\rm{erf(x)}$ is the error function.

Considering the difference between the energy reconstruction of two different kinds of experiments, we also introduce an extra parameter to incorporate a possible systematic uncertainty in the analysis. In the fit we rescale all the energies of the \phasetwo data by a factor $f$ and add a corresponding contribution $(f-1)^2/\sigma_f$ to the log-likelihood $-2 \ln \mathcal{L}$. We assume $\sigma_f$ to be 19\%, which equals the systematic uncertainty of the energy scale of \phasetwo \cite{Aharonian:2016ria}.

Following Ref.~\cite{TheFermi-LAT:2016zue}, the ALP hypothesis is evaluated by a likelihood ratio test. The maximal likelihoods under the null and ALP hypothesis are denoted by $\mathcal{L}(\mu_0|D)$ and $\mathcal{L}(\mu_{95}|D)$, respectively, where $\mu$ is the expected photon spectrum with the best fit nuisance parameters, $\mu_0$ ($mu_{95}$) is the best fit scenario without ALP (with ALP in the 0.95 quantile), respectively, and $D$ is the observed data. For each set in $(m_a, g_{a\gamma})$ plane, the adopted ICMF realization is the one among 100 realizations that corresponds to the 0.95 quantile of the likelihood distribution(the quantile of the best fit scenario corresponds to 1).

In order to test the ALP hypothesis, the probability distribution of the test statistic $TS \equiv -2 \ln (\mathcal{L}(\mu_0|D) /\mathcal{L}(\mu_{95}|D)) $ is required. Note that the relation between the spectral irregularities and ALP parameters is non-linear. Moreover only the ALP hypothesis depends on the ICMF realizations, while the null hypothesis is not. Therefore the commonly used Wilks' theorem \cite{wilks1938large} is not valid in this case. Instead, a Monte-Carlo method is needed to derive the TS distribution.

\begin{figure}[htb]
\includegraphics[width=0.50\textwidth]{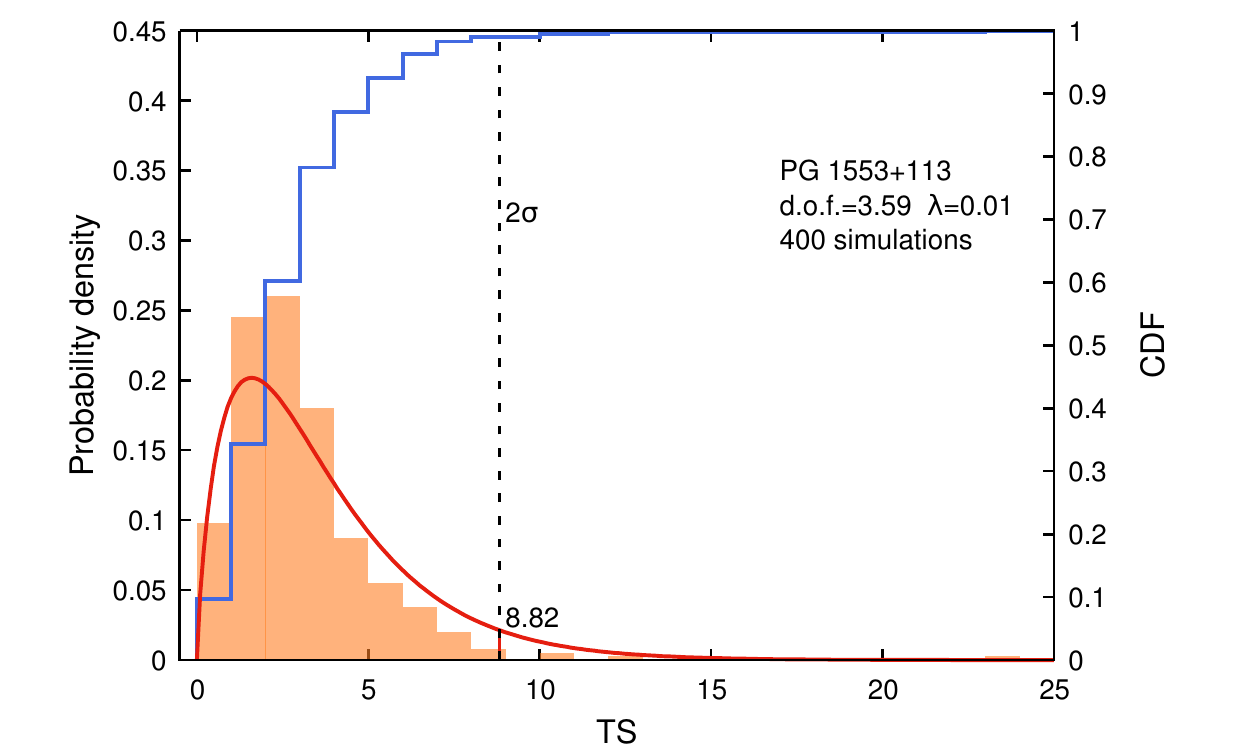}
\caption{TS distribution of \pg for the FRV08 model.
The red line represents the fitted non-central $\chi^2$ distribution with d.o.f.=3.59 and $\lambda =0.01$.
The blue line represents the cumulative probability function of TS distribution.}
\label{fig:ts}
\end{figure}

400 sets of mock data for each source are generated in pseudo-experiments that are realized by Gaussian samplings \cite{Liang:2018mqm}. For the sampling, the mean values are taken to be the best-fit fluxes under the null hypothesis; the standard deviations are taken to be the errors of the experimental data.
Then we calculate the TS value in the fit for each mock data set and derive the TS distribution.
As an example, the TS distribution of \pg for the FRV08 model is shown in Fig.~\ref{fig:ts}.
This TS distribution corresponds to a non-central $\chi^2$ distribution, where the
degree of freedom is 3.59 and non-centrality of the distribution $\lambda$ is 0.01.
The threshold of TS distribution at 95\% C.L. is found to be 8.82 and is used to
set the constraint on the ALP parameter space.

\section{Results}
\label{section_results}

\subsection{FRV08 model}
\label{section_fran_results}

\begin{figure*}[b]
\centering
\includegraphics[width=\textwidth]{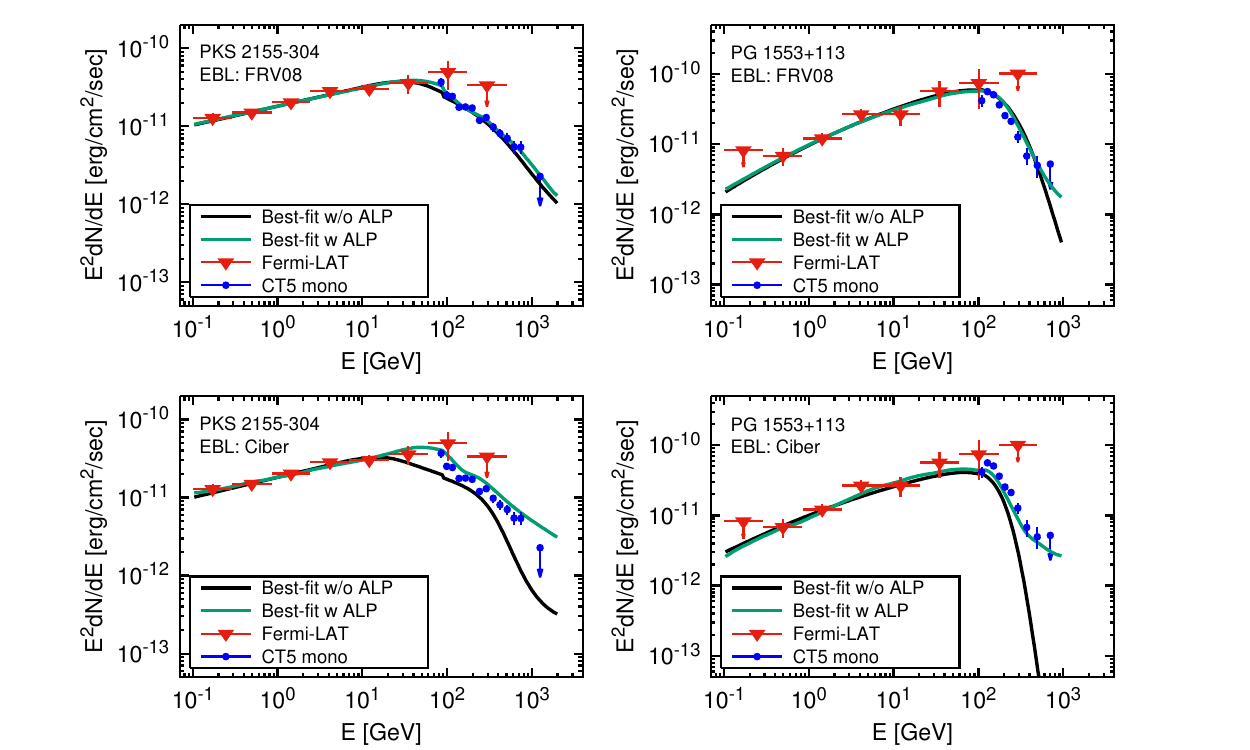}
\caption{Best-fit $\gamma-$ray spectra of \pks (left panels) and \pg (right panels). The green and black lines represent the results under the null and ALP hypothesis, respectively. The top and bottom panels represent the results for the FRV08 and Ciber EBL models, respectively. The experimental data include the results from Fermi-LAT and \phasetwo \cite{Aharonian:2016ria}.}
\label{fig:dnde}
\end{figure*}

In this section, we investigate the implication of ALP for the FRV08 EBL model. The best fit spectra under the null and ALP hypothesis for the two selected sources are shown in Fig.~\ref{fig:dnde}. It can be seen that the null hypothesis well fit the data.
The values of the best fit reduced $\chi^2$ are shown in the Table.~\ref{tab:result}.

\begin{table}[h]
\centering
\caption{The best fit $\chi^2$ and rescale factors for two sources in the two EBL models. Under the ALP hypothesis, the best fit ALP parameters $(m_a, g_{a\gamma})$ in units of $(\rm{neV}, 10^{-10}~\rm{GeV}^{-1})$} and the effective degrees of freedom of the TS distribution are also listed.
\begin{tabular*}{0.5\textwidth}{c c c c c}
    Sources&\multicolumn{2}{c} {\pks} &\multicolumn{2}{c} {\pg} \\\hline
    EBL models&FRV08 & Ciber & FRV08 & Ciber\\\hline \hline
    Best fit reduced $\chi^2$\\ w/o ALP & 22.27/16 &42.45/16 &12.95/11&28.46/11 \\\hline
    Best fit rescale \\ factor w/o ALP & 0.96 & 0.81& 1.12&0.81\\\hline \hline
    Best fit $\chi^2$ \\ w ALP & 16.31& 16.95&10.20&10.51 \\\hline
    Best fit rescale \\ factor w ALP & 1.00 & 1.12& 1.09& 0.89\\\hline
    Best fit ALP \\parameter sets & 251.19,1.58& 15.85,10&6.31,0.25&15.85,3.16 \\\hline
    Effective d.o.f of \\TS distribution & 3.98& 3.98 &3.52&2.30\\\hline
\label{tab:result}
\end{tabular*}
\end{table}

\begin{figure*}[!htbp]
\includegraphics[width=0.45\textwidth]{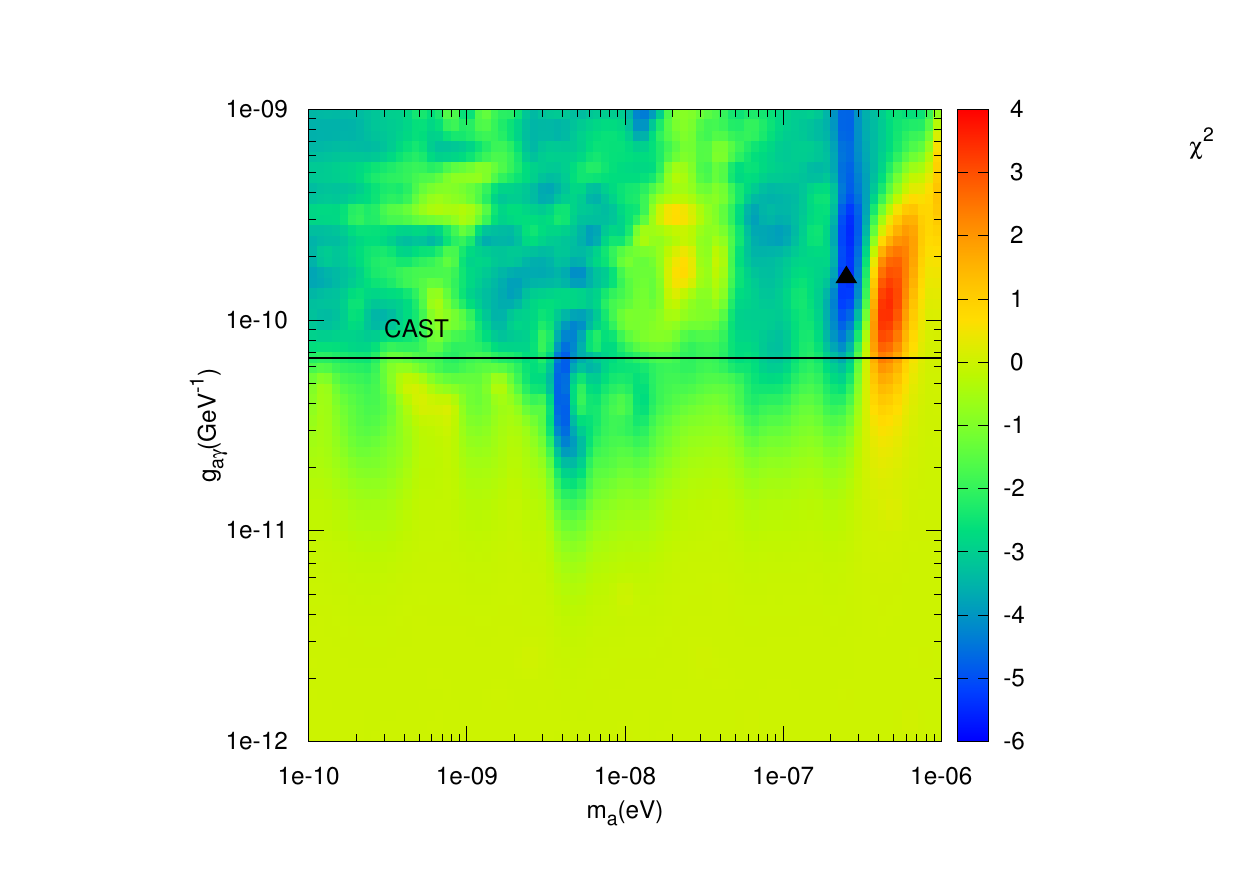}
\includegraphics[width=0.45\textwidth]{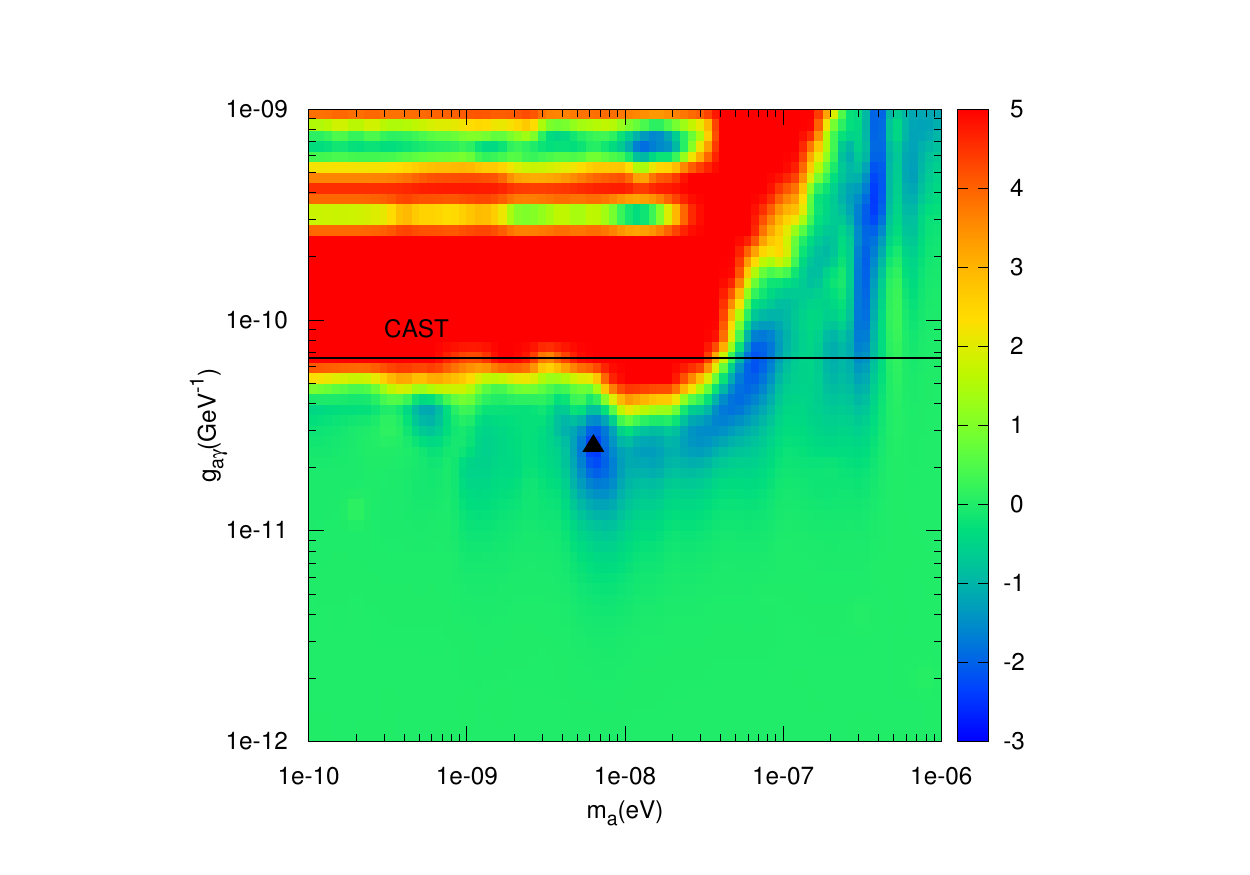}
\caption{$\Delta \chi^2\equiv \chi^2_{\rm{ALP}}-\chi^2_{\rm{null}}$ maps of \pks (left panel) and \pg (right panel). The triangle symbols represent the best fit parameters.}
\label{fig:304_delta_chi2}
\end{figure*}

Compared with the null hypothesis, the ALP-photon oscillation may reduce the EBL attenuation effect at energies above $\sim \mathcal{O}(10^2)$ GeV.
Therefore, the corresponding $\gamma$-ray spectra in this energy region may significantly deviate from the experimental data. The maps of $\Delta \chi^2\equiv \chi^2_{\rm{ALP}}-\chi^2_{\rm{null}}$ in the ($m_a, g_{a\gamma}$) plane for the two sources are shown in Fig.~\ref{fig:304_delta_chi2}. The boundaries of the excluded parameter regions can be derived by requiring $\chi^2=\chi^2_{\rm{best}}+ \chi^2_{\rm{th}}$, where $\chi^2_{\rm{best}}$ is the best-fit $\chi^2$ under the ALP hypothesis. $\chi^2_{\rm{th}}$ depending on the confidence level is taken to be the corresponding threshold of the TS distribution. Note that the probability distributions of TS with the ALP
and null hypothesis are assumed to be same here \cite{TheFermi-LAT:2016zue}. For instance, $\chi^2_{\rm{th}}$ at 95\%C.L. for \pg is taken to be 8.82.

We show the 95\% C.L. excluded contour for \pg in Fig.~\ref{fig:304_delta_chi2}. Considering the constraint from CAST, we find that the 95\% limit from \pg on the ALP-photon coupling is $g_{a\gamma} \lesssim 5 \times 10^{-11}~\rm GeV^{-1}$ in the ALP mass range of $\sim 9 ~ {\rm neV} < m_{a} < 16 ~ {\rm neV}$. For \pks, the 95\% C.L. contour is found to be above the CAST limit and is not shown here.

\begin{figure}[htb]
\includegraphics[width=0.55\textwidth]{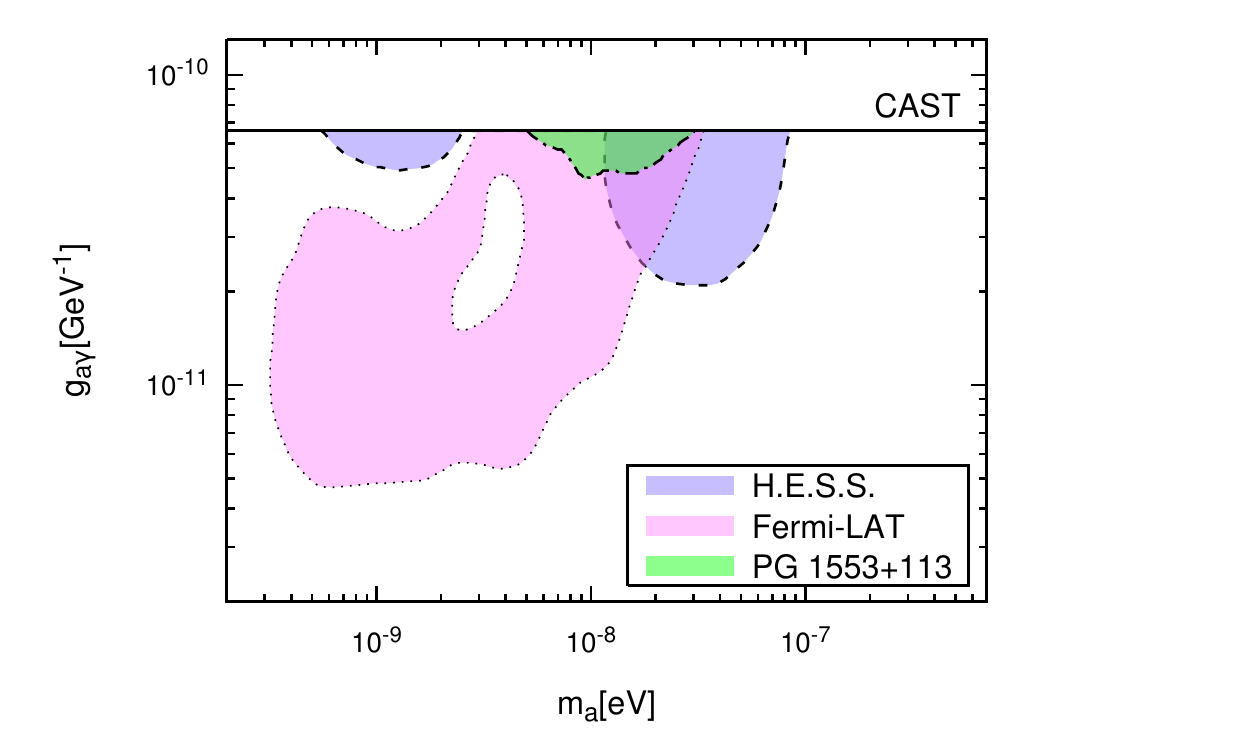}
\caption{The 95\% C.L. excluded regions in the ($m_a, g_{a\gamma}$) plane.
The green region represents our result for \pg with \(\sigma_B=3~\mu\rm G\) and the FRV08 model.
For comparison, the constraints from the CAST~\cite{Anastassopoulos:2017ftl}, Fermi-LAT observation of NGC ~1275\cite{TheFermi-LAT:2016zue} and \hess observation of \pks~\cite{Abramowski:2013oea} are also shown.}
\label{fig:FRV08_con}
\end{figure}

For comparison, the limits from Fermi-LAT observation of NGC 1275~\cite{TheFermi-LAT:2016zue})
and \hess \cite{Abramowski:2013oea} observation of \pks are also shown. The limit set by the Fermi-LAT collaboration is derived from the fit to its measured spectrum of NGC 1275. Compared with the experimental data used in this analysis, the NGC 1275 data contain more data points with narrow energy bins below $\sim 300$GeV. It is expected that the deviations from the data caused by the ALP-photon oscillation would be more significant at low energies.
Therefore, the Fermi-LAT analysis has excluded a large parameter region at low ALP masses $\sim \mathcal{O}(1)$ neV, which correspond to low critical energies for the ALP-photon oscillation.
The \pks analysis of the \hess collaboration focuses on the spectral irregularities induced by the ALP-photon oscillation in the variations of neighboring energy bins and provides a stricter limit in comparison with our result for \pks.

\subsection{Ciber model}
\label{section_ciber_results}

The implication of ALP would change for the Ciber EBL model. The best fit spectra of the null and ALP hypotheses for the two sources are shown in the bottom panels of Fig.~\ref{fig:dnde}.
Compared with the FRV08 model, the excess at $\sim \mu$m in the Ciber model induce an additional attenuation effect above $\sim 300~$GeV and lead to more significant deviations from the data, which can be seen from Table.~\ref{tab:result}. The ALP-photon oscillation may compensate this additional attenuation effect and improve the fit to the data.
This improvement method has been discussed for the Fermi-LAT and \hess observations of two sources H2356-309 (z=0.165) and 1ES1101-232 (z=0.186) through the $\chi^2$ fit in Ref.~\cite{Kohri:2017ljt}.

\begin{figure}[htb]
\includegraphics[width=0.55\textwidth]{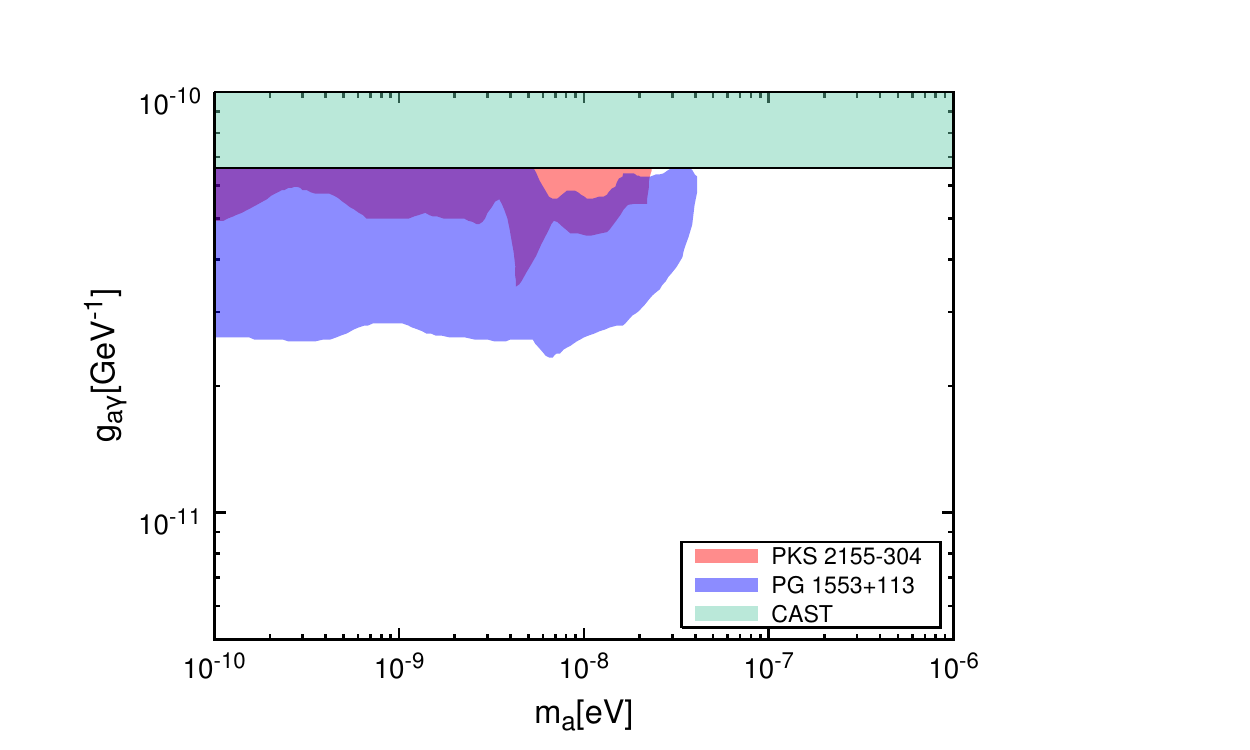}
\caption{Favored ALP parameter region where the fit to the \pks and \pg observations can be improved at 95\%C.L..}
\label{fig:ciber_con}
\end{figure}

We show the improvement regions at 95\% C.L. for \pks and \pg in Fig.~\ref{fig:ciber_con}.
The favored ALP parameter region for \pg is almost a rectangular region with $g_{a\gamma}\gtrsim 2.6\times 10^{-11}~\rm{GeV}^{-1}$ and $ m_a \lesssim 10~{\rm neV}$.
The favored region for \pks is about $g_{a\gamma}\gtrsim 5\times 10^{-11}~\rm{GeV}^{-1}$ and $ m_a \lesssim 15~{\rm neV}$.
Compared with the favored region derived in Ref.~\cite{Kohri:2017ljt} ($g_{a\gamma}\gtrsim 2\times 10^{-11}~\rm{GeV}^{-1}$ for $1~\rm{neV}\lesssim m_a \lesssim 40~\rm{neV}$), there is no lower boundary on $m_a$ in our results. This is because that the ALP-photon oscillation effect in the extragalactic space is neglected in this analysis.

\section{Conclusion}
\label{section_Con}

In this work, we investigated the ALP implication in the Fermi-LAT and \phasetwo gamma-ray observations of two sources \pg and \pks.
Two EBL models are considered in the analysis. We found that the best fit spectra under the null hypothesis can well fit the experimental data for the FRV08 EBL model and set constraints on the ALP parameter region in the ($m_a,g_{a\gamma}$) plane. For $\sigma_B=3~\mu$G, the constraint on $g_{a\gamma}$ at 95\% C.L. is $g_{a\gamma}\lesssim 5\times 10^{-11} ~\rm{GeV}^{-1}$ for an ALP mass between $9$ and $16$ neV.
On the other hand, we found that the ALP-photon oscillation would improve the fit to the \pks and \pg observations for the Ciber model with an excess at $\sim 1 \mu$m. The favored parameter region is given.

In future, Cherenkov high energy gamma-ray telescopes will provide more accurate results. The large ground-based telescopes, such as CTA\cite{Acharya:2013sxa} and LHAASO\cite{Cao:2010zz}, will measure the spectra of extragalatic gamma-ray sources at very high energies. Combined with these results, it is possible to search for the spectral regularities induced by the ALP-photon oscillation and accurately investigate the ALP implication in high energy astrophysical processes.

\section*{Acknowledgments}
This work is supported by the National Key R\&D Program of China
(No. 2016YFA0400200), the National Natural Science Foundation of China
(Nos. U1738209 and 11851303).

\bibliography{reference}

\begin{thebibliography}{72}
\expandafter\ifx\csname natexlab\endcsname\relax\def\natexlab#1{#1}\fi
\expandafter\ifx\csname bibnamefont\endcsname\relax
  \def\bibnamefont#1{#1}\fi
\expandafter\ifx\csname bibfnamefont\endcsname\relax
  \def\bibfnamefont#1{#1}\fi
\expandafter\ifx\csname citenamefont\endcsname\relax
  \def\citenamefont#1{#1}\fi
\expandafter\ifx\csname url\endcsname\relax
  \def\url#1{\texttt{#1}}\fi
\expandafter\ifx\csname urlprefix\endcsname\relax\def\urlprefix{URL }\fi
\providecommand{\bibinfo}[2]{#2}
\providecommand{\eprint}[2][]{\url{#2}}

\bibitem[{\citenamefont{Peccei and Quinn}(1977)}]{Peccei:1977hh}
\bibinfo{author}{\bibfnamefont{R.~D.} \bibnamefont{Peccei}} \bibnamefont{and}
  \bibinfo{author}{\bibfnamefont{H.~R.} \bibnamefont{Quinn}},
  \bibinfo{journal}{Phys. Rev. Lett.} \textbf{\bibinfo{volume}{38}},
  \bibinfo{pages}{1440} (\bibinfo{year}{1977}), \bibinfo{note}{[,328(1977)]}.

\bibitem[{\citenamefont{Weinberg}(1978)}]{Weinberg:1977ma}
\bibinfo{author}{\bibfnamefont{S.}~\bibnamefont{Weinberg}},
  \bibinfo{journal}{Phys. Rev. Lett.} \textbf{\bibinfo{volume}{40}},
  \bibinfo{pages}{223} (\bibinfo{year}{1978}).

\bibitem[{\citenamefont{Peccei}(2008)}]{Peccei:2006as}
\bibinfo{author}{\bibfnamefont{R.~D.} \bibnamefont{Peccei}},
  \bibinfo{journal}{Lect. Notes Phys.} \textbf{\bibinfo{volume}{741}},
  \bibinfo{pages}{3} (\bibinfo{year}{2008}), \bibinfo{note}{[,3(2006)]},
  \eprint{hep-ph/0607268}.

\bibitem[{\citenamefont{Cicoli et~al.}(2012)\citenamefont{Cicoli, Goodsell, and
  Ringwald}}]{Cicoli:2012sz}
\bibinfo{author}{\bibfnamefont{M.}~\bibnamefont{Cicoli}},
  \bibinfo{author}{\bibfnamefont{M.}~\bibnamefont{Goodsell}}, \bibnamefont{and}
  \bibinfo{author}{\bibfnamefont{A.}~\bibnamefont{Ringwald}},
  \bibinfo{journal}{JHEP} \textbf{\bibinfo{volume}{10}}, \bibinfo{pages}{146}
  (\bibinfo{year}{2012}), \eprint{1206.0819}.

\bibitem[{\citenamefont{Di~Lella et~al.}(2000)\citenamefont{Di~Lella,
  Pilaftsis, Raffelt, and Zioutas}}]{DiLella:2000dn}
\bibinfo{author}{\bibfnamefont{L.}~\bibnamefont{Di~Lella}},
  \bibinfo{author}{\bibfnamefont{A.}~\bibnamefont{Pilaftsis}},
  \bibinfo{author}{\bibfnamefont{G.}~\bibnamefont{Raffelt}}, \bibnamefont{and}
  \bibinfo{author}{\bibfnamefont{K.}~\bibnamefont{Zioutas}},
  \bibinfo{journal}{Phys. Rev.} \textbf{\bibinfo{volume}{D62}},
  \bibinfo{pages}{125011} (\bibinfo{year}{2000}), \eprint{hep-ph/0006327}.

\bibitem[{\citenamefont{Sikivie}(1983)}]{sikivie1983experimental}
\bibinfo{author}{\bibfnamefont{P.}~\bibnamefont{Sikivie}},
  \bibinfo{journal}{Physical Review Letters} \textbf{\bibinfo{volume}{51}},
  \bibinfo{pages}{1415} (\bibinfo{year}{1983}).

\bibitem[{\citenamefont{Asztalos et~al.}(2010)\citenamefont{Asztalos, Carosi,
  Hagmann, Kinion, Van~Bibber, Hotz, Rosenberg, Rybka, Hoskins, Hwang
  et~al.}}]{asztalos2010squid}
\bibinfo{author}{\bibfnamefont{S.~J.} \bibnamefont{Asztalos}},
  \bibinfo{author}{\bibfnamefont{G.}~\bibnamefont{Carosi}},
  \bibinfo{author}{\bibfnamefont{C.}~\bibnamefont{Hagmann}},
  \bibinfo{author}{\bibfnamefont{D.}~\bibnamefont{Kinion}},
  \bibinfo{author}{\bibfnamefont{K.}~\bibnamefont{Van~Bibber}},
  \bibinfo{author}{\bibfnamefont{M.}~\bibnamefont{Hotz}},
  \bibinfo{author}{\bibfnamefont{L.}~\bibnamefont{Rosenberg}},
  \bibinfo{author}{\bibfnamefont{G.}~\bibnamefont{Rybka}},
  \bibinfo{author}{\bibfnamefont{J.}~\bibnamefont{Hoskins}},
  \bibinfo{author}{\bibfnamefont{J.}~\bibnamefont{Hwang}},
  \bibnamefont{et~al.}, \bibinfo{journal}{Physical review letters}
  \textbf{\bibinfo{volume}{104}}, \bibinfo{pages}{041301}
  (\bibinfo{year}{2010}).

\bibitem[{\citenamefont{Carosi et~al.}(2013)\citenamefont{Carosi, Friedland,
  Giannotti, Pivovaroff, Ruz, and Vogel}}]{carosi2013probing}
\bibinfo{author}{\bibfnamefont{G.}~\bibnamefont{Carosi}},
  \bibinfo{author}{\bibfnamefont{A.}~\bibnamefont{Friedland}},
  \bibinfo{author}{\bibfnamefont{M.}~\bibnamefont{Giannotti}},
  \bibinfo{author}{\bibfnamefont{M.}~\bibnamefont{Pivovaroff}},
  \bibinfo{author}{\bibfnamefont{J.}~\bibnamefont{Ruz}}, \bibnamefont{and}
  \bibinfo{author}{\bibfnamefont{J.}~\bibnamefont{Vogel}},
  \bibinfo{journal}{arXiv preprint arXiv:1309.7035}  (\bibinfo{year}{2013}).

\bibitem[{\citenamefont{Graham et~al.}(2015)\citenamefont{Graham, Irastorza,
  Lamoreaux, Lindner, and van Bibber}}]{Graham:2015ouw}
\bibinfo{author}{\bibfnamefont{P.~W.} \bibnamefont{Graham}},
  \bibinfo{author}{\bibfnamefont{I.~G.} \bibnamefont{Irastorza}},
  \bibinfo{author}{\bibfnamefont{S.~K.} \bibnamefont{Lamoreaux}},
  \bibinfo{author}{\bibfnamefont{A.}~\bibnamefont{Lindner}}, \bibnamefont{and}
  \bibinfo{author}{\bibfnamefont{K.~A.} \bibnamefont{van Bibber}},
  \bibinfo{journal}{Ann. Rev. Nucl. Part. Sci.} \textbf{\bibinfo{volume}{65}},
  \bibinfo{pages}{485} (\bibinfo{year}{2015}), \eprint{1602.00039}.

\bibitem[{\citenamefont{Majumdar et~al.}(2017)\citenamefont{Majumdar, Calore,
  and Horns}}]{Majumdar:2017vcx}
\bibinfo{author}{\bibfnamefont{J.}~\bibnamefont{Majumdar}},
  \bibinfo{author}{\bibfnamefont{F.}~\bibnamefont{Calore}}, \bibnamefont{and}
  \bibinfo{author}{\bibfnamefont{D.}~\bibnamefont{Horns}},
  \bibinfo{journal}{PoS} \textbf{\bibinfo{volume}{IFS2017}},
  \bibinfo{pages}{168} (\bibinfo{year}{2017}), \eprint{1711.08723}.

\bibitem[{\citenamefont{Galanti and Roncadelli}(2018)}]{Galanti:2018myb}
\bibinfo{author}{\bibfnamefont{G.}~\bibnamefont{Galanti}} \bibnamefont{and}
  \bibinfo{author}{\bibfnamefont{M.}~\bibnamefont{Roncadelli}},
  \bibinfo{journal}{JHEAp} \textbf{\bibinfo{volume}{20}}, \bibinfo{pages}{1}
  (\bibinfo{year}{2018}), \eprint{1805.12055}.

\bibitem[{\citenamefont{Anastassopoulos
  et~al.}(2017)}]{Anastassopoulos:2017ftl}
\bibinfo{author}{\bibfnamefont{V.}~\bibnamefont{Anastassopoulos}}
  \bibnamefont{et~al.} (\bibinfo{collaboration}{CAST}),
  \bibinfo{journal}{Nature Phys.} \textbf{\bibinfo{volume}{13}},
  \bibinfo{pages}{584} (\bibinfo{year}{2017}), \eprint{1705.02290}.

\bibitem[{\citenamefont{Raffelt}(1990)}]{Raffelt:1990yz}
\bibinfo{author}{\bibfnamefont{G.~G.} \bibnamefont{Raffelt}},
  \bibinfo{journal}{Phys. Rept.} \textbf{\bibinfo{volume}{198}},
  \bibinfo{pages}{1} (\bibinfo{year}{1990}).

\bibitem[{\citenamefont{Raffelt and Stodolsky}(1988)}]{Raffelt:1987im}
\bibinfo{author}{\bibfnamefont{G.}~\bibnamefont{Raffelt}} \bibnamefont{and}
  \bibinfo{author}{\bibfnamefont{L.}~\bibnamefont{Stodolsky}},
  \bibinfo{journal}{Phys. Rev.} \textbf{\bibinfo{volume}{D37}},
  \bibinfo{pages}{1237} (\bibinfo{year}{1988}).

\bibitem[{\citenamefont{Mirizzi and Montanino}(2009)}]{Mirizzi:2009aj}
\bibinfo{author}{\bibfnamefont{A.}~\bibnamefont{Mirizzi}} \bibnamefont{and}
  \bibinfo{author}{\bibfnamefont{D.}~\bibnamefont{Montanino}},
  \bibinfo{journal}{JCAP} \textbf{\bibinfo{volume}{0912}}, \bibinfo{pages}{004}
  (\bibinfo{year}{2009}), \eprint{0911.0015}.

\bibitem[{\citenamefont{De~Angelis et~al.}(2011)\citenamefont{De~Angelis,
  Galanti, and Roncadelli}}]{DeAngelis:2011id}
\bibinfo{author}{\bibfnamefont{A.}~\bibnamefont{De~Angelis}},
  \bibinfo{author}{\bibfnamefont{G.}~\bibnamefont{Galanti}}, \bibnamefont{and}
  \bibinfo{author}{\bibfnamefont{M.}~\bibnamefont{Roncadelli}},
  \bibinfo{journal}{Phys. Rev.} \textbf{\bibinfo{volume}{D84}},
  \bibinfo{pages}{105030} (\bibinfo{year}{2011}), \bibinfo{note}{[Erratum:
  Phys. Rev.D87,no.10,109903(2013)]}, \eprint{1106.1132}.

\bibitem[{\citenamefont{Belikov et~al.}(2011)\citenamefont{Belikov, Goodenough,
  and Hooper}}]{Belikov:2010ma}
\bibinfo{author}{\bibfnamefont{A.~V.} \bibnamefont{Belikov}},
  \bibinfo{author}{\bibfnamefont{L.}~\bibnamefont{Goodenough}},
  \bibnamefont{and} \bibinfo{author}{\bibfnamefont{D.}~\bibnamefont{Hooper}},
  \bibinfo{journal}{Phys. Rev.} \textbf{\bibinfo{volume}{D83}},
  \bibinfo{pages}{063005} (\bibinfo{year}{2011}), \eprint{1007.4862}.

\bibitem[{\citenamefont{Abramowski et~al.}(2013)}]{Abramowski:2013oea}
\bibinfo{author}{\bibfnamefont{A.}~\bibnamefont{Abramowski}}
  \bibnamefont{et~al.} (\bibinfo{collaboration}{H.E.S.S.}),
  \bibinfo{journal}{Phys. Rev.} \textbf{\bibinfo{volume}{D88}},
  \bibinfo{pages}{102003} (\bibinfo{year}{2013}), \eprint{1311.3148}.

\bibitem[{\citenamefont{Reesman and Walker}(2014)}]{Reesman:2014ova}
\bibinfo{author}{\bibfnamefont{R.}~\bibnamefont{Reesman}} \bibnamefont{and}
  \bibinfo{author}{\bibfnamefont{T.~P.} \bibnamefont{Walker}},
  \bibinfo{journal}{JCAP} \textbf{\bibinfo{volume}{1408}}, \bibinfo{pages}{021}
  (\bibinfo{year}{2014}), \eprint{1402.2533}.

\bibitem[{\citenamefont{Payez et~al.}(2015)\citenamefont{Payez, Evoli, Fischer,
  Giannotti, Mirizzi, and Ringwald}}]{Payez:2014xsa}
\bibinfo{author}{\bibfnamefont{A.}~\bibnamefont{Payez}},
  \bibinfo{author}{\bibfnamefont{C.}~\bibnamefont{Evoli}},
  \bibinfo{author}{\bibfnamefont{T.}~\bibnamefont{Fischer}},
  \bibinfo{author}{\bibfnamefont{M.}~\bibnamefont{Giannotti}},
  \bibinfo{author}{\bibfnamefont{A.}~\bibnamefont{Mirizzi}}, \bibnamefont{and}
  \bibinfo{author}{\bibfnamefont{A.}~\bibnamefont{Ringwald}},
  \bibinfo{journal}{JCAP} \textbf{\bibinfo{volume}{1502}}, \bibinfo{pages}{006}
  (\bibinfo{year}{2015}), \eprint{1410.3747}.

\bibitem[{\citenamefont{Berenji et~al.}(2016)\citenamefont{Berenji, Gaskins,
  and Meyer}}]{Berenji:2016jji}
\bibinfo{author}{\bibfnamefont{B.}~\bibnamefont{Berenji}},
  \bibinfo{author}{\bibfnamefont{J.}~\bibnamefont{Gaskins}}, \bibnamefont{and}
  \bibinfo{author}{\bibfnamefont{M.}~\bibnamefont{Meyer}},
  \bibinfo{journal}{Phys. Rev.} \textbf{\bibinfo{volume}{D93}},
  \bibinfo{pages}{045019} (\bibinfo{year}{2016}), \eprint{1602.00091}.

\bibitem[{\citenamefont{Ajello et~al.}(2016)}]{TheFermi-LAT:2016zue}
\bibinfo{author}{\bibfnamefont{M.}~\bibnamefont{Ajello}} \bibnamefont{et~al.}
  (\bibinfo{collaboration}{Fermi-LAT}), \bibinfo{journal}{Phys. Rev. Lett.}
  \textbf{\bibinfo{volume}{116}}, \bibinfo{pages}{161101}
  (\bibinfo{year}{2016}), \eprint{1603.06978}.

\bibitem[{\citenamefont{Meyer et~al.}(2017)\citenamefont{Meyer, Giannotti,
  Mirizzi, Conrad, and Sánchez-Conde}}]{Meyer:2016wrm}
\bibinfo{author}{\bibfnamefont{M.}~\bibnamefont{Meyer}},
  \bibinfo{author}{\bibfnamefont{M.}~\bibnamefont{Giannotti}},
  \bibinfo{author}{\bibfnamefont{A.}~\bibnamefont{Mirizzi}},
  \bibinfo{author}{\bibfnamefont{J.}~\bibnamefont{Conrad}}, \bibnamefont{and}
  \bibinfo{author}{\bibfnamefont{M.~A.} \bibnamefont{Sánchez-Conde}},
  \bibinfo{journal}{Phys. Rev. Lett.} \textbf{\bibinfo{volume}{118}},
  \bibinfo{pages}{011103} (\bibinfo{year}{2017}), \eprint{1609.02350}.

\bibitem[{\citenamefont{Liang et~al.}(2018)\citenamefont{Liang, Zhang, Xia,
  Feng, Yuan, and Fan}}]{Liang:2018mqm}
\bibinfo{author}{\bibfnamefont{Y.-F.} \bibnamefont{Liang}},
  \bibinfo{author}{\bibfnamefont{C.}~\bibnamefont{Zhang}},
  \bibinfo{author}{\bibfnamefont{Z.-Q.} \bibnamefont{Xia}},
  \bibinfo{author}{\bibfnamefont{L.}~\bibnamefont{Feng}},
  \bibinfo{author}{\bibfnamefont{Q.}~\bibnamefont{Yuan}}, \bibnamefont{and}
  \bibinfo{author}{\bibfnamefont{Y.-Z.} \bibnamefont{Fan}}
  (\bibinfo{year}{2018}), \eprint{1804.07186}.

\bibitem[{\citenamefont{Zhang et~al.}(2018)\citenamefont{Zhang, Liang, Li,
  Liao, Feng, Yuan, Fan, and Ren}}]{Zhang:2018wpc}
\bibinfo{author}{\bibfnamefont{C.}~\bibnamefont{Zhang}},
  \bibinfo{author}{\bibfnamefont{Y.-F.} \bibnamefont{Liang}},
  \bibinfo{author}{\bibfnamefont{S.}~\bibnamefont{Li}},
  \bibinfo{author}{\bibfnamefont{N.-H.} \bibnamefont{Liao}},
  \bibinfo{author}{\bibfnamefont{L.}~\bibnamefont{Feng}},
  \bibinfo{author}{\bibfnamefont{Q.}~\bibnamefont{Yuan}},
  \bibinfo{author}{\bibfnamefont{Y.-Z.} \bibnamefont{Fan}}, \bibnamefont{and}
  \bibinfo{author}{\bibfnamefont{Z.-Z.} \bibnamefont{Ren}},
  \bibinfo{journal}{Phys. Rev.} \textbf{\bibinfo{volume}{D97}},
  \bibinfo{pages}{063009} (\bibinfo{year}{2018}), \eprint{1802.08420}.

\bibitem[{\citenamefont{Libanov and Troitsky}(2019)}]{Libanov:2019fzq}
\bibinfo{author}{\bibfnamefont{M.}~\bibnamefont{Libanov}} \bibnamefont{and}
  \bibinfo{author}{\bibfnamefont{S.}~\bibnamefont{Troitsky}}
  (\bibinfo{year}{2019}), \eprint{1908.03084}.

\bibitem[{\citenamefont{Long et~al.}(2019)\citenamefont{Long, Lin, Tam, and
  Zhu}}]{Long:2019nrz}
\bibinfo{author}{\bibfnamefont{G.~B.} \bibnamefont{Long}},
  \bibinfo{author}{\bibfnamefont{W.~P.} \bibnamefont{Lin}},
  \bibinfo{author}{\bibfnamefont{P.~H.~T.} \bibnamefont{Tam}},
  \bibnamefont{and} \bibinfo{author}{\bibfnamefont{W.~S.} \bibnamefont{Zhu}}
  (\bibinfo{year}{2019}), \eprint{1912.05309}.

\bibitem[{\citenamefont{Troitsky}(2016)}]{Troitsky:2015nxa}
\bibinfo{author}{\bibfnamefont{S.}~\bibnamefont{Troitsky}},
  \bibinfo{journal}{Phys. Rev.} \textbf{\bibinfo{volume}{D93}},
  \bibinfo{pages}{045014} (\bibinfo{year}{2016}), \eprint{1507.08640}.

\bibitem[{\citenamefont{Csaki et~al.}(2003)\citenamefont{Csaki, Kaloper,
  Peloso, and Terning}}]{Csaki:2003ef}
\bibinfo{author}{\bibfnamefont{C.}~\bibnamefont{Csaki}},
  \bibinfo{author}{\bibfnamefont{N.}~\bibnamefont{Kaloper}},
  \bibinfo{author}{\bibfnamefont{M.}~\bibnamefont{Peloso}}, \bibnamefont{and}
  \bibinfo{author}{\bibfnamefont{J.}~\bibnamefont{Terning}},
  \bibinfo{journal}{JCAP} \textbf{\bibinfo{volume}{0305}}, \bibinfo{pages}{005}
  (\bibinfo{year}{2003}), \eprint{hep-ph/0302030}.

\bibitem[{\citenamefont{De~Angelis et~al.}(2007)\citenamefont{De~Angelis,
  Roncadelli, and Mansutti}}]{DeAngelis:2007dqd}
\bibinfo{author}{\bibfnamefont{A.}~\bibnamefont{De~Angelis}},
  \bibinfo{author}{\bibfnamefont{M.}~\bibnamefont{Roncadelli}},
  \bibnamefont{and} \bibinfo{author}{\bibfnamefont{O.}~\bibnamefont{Mansutti}},
  \bibinfo{journal}{Phys. Rev.} \textbf{\bibinfo{volume}{D76}},
  \bibinfo{pages}{121301} (\bibinfo{year}{2007}), \eprint{0707.4312}.

\bibitem[{\citenamefont{Simet et~al.}(2008)\citenamefont{Simet, Hooper, and
  Serpico}}]{Simet:2007sa}
\bibinfo{author}{\bibfnamefont{M.}~\bibnamefont{Simet}},
  \bibinfo{author}{\bibfnamefont{D.}~\bibnamefont{Hooper}}, \bibnamefont{and}
  \bibinfo{author}{\bibfnamefont{P.~D.} \bibnamefont{Serpico}},
  \bibinfo{journal}{Phys. Rev.} \textbf{\bibinfo{volume}{D77}},
  \bibinfo{pages}{063001} (\bibinfo{year}{2008}), \eprint{0712.2825}.

\bibitem[{\citenamefont{Fairbairn et~al.}(2011)\citenamefont{Fairbairn, Rashba,
  and Troitsky}}]{Fairbairn:2009zi}
\bibinfo{author}{\bibfnamefont{M.}~\bibnamefont{Fairbairn}},
  \bibinfo{author}{\bibfnamefont{T.}~\bibnamefont{Rashba}}, \bibnamefont{and}
  \bibinfo{author}{\bibfnamefont{S.~V.} \bibnamefont{Troitsky}},
  \bibinfo{journal}{Phys. Rev.} \textbf{\bibinfo{volume}{D84}},
  \bibinfo{pages}{125019} (\bibinfo{year}{2011}), \eprint{0901.4085}.

\bibitem[{\citenamefont{Meyer et~al.}(2013)\citenamefont{Meyer, Horns, and
  Raue}}]{Meyer:2013pny}
\bibinfo{author}{\bibfnamefont{M.}~\bibnamefont{Meyer}},
  \bibinfo{author}{\bibfnamefont{D.}~\bibnamefont{Horns}}, \bibnamefont{and}
  \bibinfo{author}{\bibfnamefont{M.}~\bibnamefont{Raue}},
  \bibinfo{journal}{Phys. Rev.} \textbf{\bibinfo{volume}{D87}},
  \bibinfo{pages}{035027} (\bibinfo{year}{2013}), \eprint{1302.1208}.

\bibitem[{\citenamefont{Dominguez
  et~al.}(2011{\natexlab{a}})\citenamefont{Dominguez, Sanchez-Conde, and
  Prada}}]{Dominguez:2011xy}
\bibinfo{author}{\bibfnamefont{A.}~\bibnamefont{Dominguez}},
  \bibinfo{author}{\bibfnamefont{M.~A.} \bibnamefont{Sanchez-Conde}},
  \bibnamefont{and} \bibinfo{author}{\bibfnamefont{F.}~\bibnamefont{Prada}},
  \bibinfo{journal}{JCAP} \textbf{\bibinfo{volume}{1111}}, \bibinfo{pages}{020}
  (\bibinfo{year}{2011}{\natexlab{a}}), \eprint{1106.1860}.

\bibitem[{\citenamefont{Mirizzi et~al.}(2007)\citenamefont{Mirizzi, Raffelt,
  and Serpico}}]{Mirizzi:2007hr}
\bibinfo{author}{\bibfnamefont{A.}~\bibnamefont{Mirizzi}},
  \bibinfo{author}{\bibfnamefont{G.~G.} \bibnamefont{Raffelt}},
  \bibnamefont{and} \bibinfo{author}{\bibfnamefont{P.~D.}
  \bibnamefont{Serpico}}, \bibinfo{journal}{Phys. Rev.}
  \textbf{\bibinfo{volume}{D76}}, \bibinfo{pages}{023001}
  (\bibinfo{year}{2007}), \eprint{0704.3044}.

\bibitem[{\citenamefont{De~Angelis et~al.}(2009)\citenamefont{De~Angelis,
  Mansutti, Persic, and Roncadelli}}]{de2009photon}
\bibinfo{author}{\bibfnamefont{A.}~\bibnamefont{De~Angelis}},
  \bibinfo{author}{\bibfnamefont{O.}~\bibnamefont{Mansutti}},
  \bibinfo{author}{\bibfnamefont{M.}~\bibnamefont{Persic}}, \bibnamefont{and}
  \bibinfo{author}{\bibfnamefont{M.}~\bibnamefont{Roncadelli}},
  \bibinfo{journal}{Monthly Notices of the Royal Astronomical Society: Letters}
  \textbf{\bibinfo{volume}{394}}, \bibinfo{pages}{L21} (\bibinfo{year}{2009}).

\bibitem[{\citenamefont{Sanchez-Conde et~al.}(2009)\citenamefont{Sanchez-Conde,
  Paneque, Bloom, Prada, and Dominguez}}]{SanchezConde:2009wu}
\bibinfo{author}{\bibfnamefont{M.~A.} \bibnamefont{Sanchez-Conde}},
  \bibinfo{author}{\bibfnamefont{D.}~\bibnamefont{Paneque}},
  \bibinfo{author}{\bibfnamefont{E.}~\bibnamefont{Bloom}},
  \bibinfo{author}{\bibfnamefont{F.}~\bibnamefont{Prada}}, \bibnamefont{and}
  \bibinfo{author}{\bibfnamefont{A.}~\bibnamefont{Dominguez}},
  \bibinfo{journal}{Phys. Rev.} \textbf{\bibinfo{volume}{D79}},
  \bibinfo{pages}{123511} (\bibinfo{year}{2009}), \eprint{0905.3270}.

\bibitem[{\citenamefont{Kohri and Kodama}(2017)}]{Kohri:2017ljt}
\bibinfo{author}{\bibfnamefont{K.}~\bibnamefont{Kohri}} \bibnamefont{and}
  \bibinfo{author}{\bibfnamefont{H.}~\bibnamefont{Kodama}},
  \bibinfo{journal}{Phys. Rev.} \textbf{\bibinfo{volume}{D96}},
  \bibinfo{pages}{051701} (\bibinfo{year}{2017}), \eprint{1704.05189}.

\bibitem[{\citenamefont{Bi et~al.}(2020)\citenamefont{Bi, Gao, Guo, Houston,
  Li, Xu, and Zhang}}]{Bi:2020ths}
\bibinfo{author}{\bibfnamefont{X.-J.} \bibnamefont{Bi}},
  \bibinfo{author}{\bibfnamefont{Y.}~\bibnamefont{Gao}},
  \bibinfo{author}{\bibfnamefont{J.}~\bibnamefont{Guo}},
  \bibinfo{author}{\bibfnamefont{N.}~\bibnamefont{Houston}},
  \bibinfo{author}{\bibfnamefont{T.}~\bibnamefont{Li}},
  \bibinfo{author}{\bibfnamefont{F.}~\bibnamefont{Xu}}, \bibnamefont{and}
  \bibinfo{author}{\bibfnamefont{X.}~\bibnamefont{Zhang}}
  (\bibinfo{year}{2020}), \eprint{2002.01796}.

\bibitem[{\citenamefont{Abdalla et~al.}(2017)}]{Aharonian:2016ria}
\bibinfo{author}{\bibfnamefont{H.}~\bibnamefont{Abdalla}} \bibnamefont{et~al.}
  (\bibinfo{collaboration}{H.E.S.S., LAT}), \bibinfo{journal}{Astron.
  Astrophys.} \textbf{\bibinfo{volume}{600}}, \bibinfo{pages}{A89}
  (\bibinfo{year}{2017}), \eprint{1612.01843}.

\bibitem[{\citenamefont{Franceschini et~al.}(2008)\citenamefont{Franceschini,
  Rodighiero, and Vaccari}}]{Franceschini:2008tp}
\bibinfo{author}{\bibfnamefont{A.}~\bibnamefont{Franceschini}},
  \bibinfo{author}{\bibfnamefont{G.}~\bibnamefont{Rodighiero}},
  \bibnamefont{and} \bibinfo{author}{\bibfnamefont{M.}~\bibnamefont{Vaccari}},
  \bibinfo{journal}{Astron. Astrophys.} \textbf{\bibinfo{volume}{487}},
  \bibinfo{pages}{837} (\bibinfo{year}{2008}), \eprint{0805.1841}.

\bibitem[{\citenamefont{Finke et~al.}(2010)\citenamefont{Finke, Razzaque, and
  Dermer}}]{Finke:2009xi}
\bibinfo{author}{\bibfnamefont{J.~D.} \bibnamefont{Finke}},
  \bibinfo{author}{\bibfnamefont{S.}~\bibnamefont{Razzaque}}, \bibnamefont{and}
  \bibinfo{author}{\bibfnamefont{C.~D.} \bibnamefont{Dermer}},
  \bibinfo{journal}{Astrophys. J.} \textbf{\bibinfo{volume}{712}},
  \bibinfo{pages}{238} (\bibinfo{year}{2010}), \eprint{0905.1115}.

\bibitem[{\citenamefont{Dominguez
  et~al.}(2011{\natexlab{b}})}]{Dominguez:2010bv}
\bibinfo{author}{\bibfnamefont{A.}~\bibnamefont{Dominguez}}
  \bibnamefont{et~al.}, \bibinfo{journal}{Mon. Not. Roy. Astron. Soc.}
  \textbf{\bibinfo{volume}{410}}, \bibinfo{pages}{2556}
  (\bibinfo{year}{2011}{\natexlab{b}}), \eprint{1007.1459}.

\bibitem[{\citenamefont{Gilmore et~al.}(2012)\citenamefont{Gilmore, Somerville,
  Primack, and Dominguez}}]{Gilmore:2011ks}
\bibinfo{author}{\bibfnamefont{R.~C.} \bibnamefont{Gilmore}},
  \bibinfo{author}{\bibfnamefont{R.~S.} \bibnamefont{Somerville}},
  \bibinfo{author}{\bibfnamefont{J.~R.} \bibnamefont{Primack}},
  \bibnamefont{and}
  \bibinfo{author}{\bibfnamefont{A.}~\bibnamefont{Dominguez}},
  \bibinfo{journal}{Mon. Not. Roy. Astron. Soc.}
  \textbf{\bibinfo{volume}{422}}, \bibinfo{pages}{3189} (\bibinfo{year}{2012}),
  \eprint{1104.0671}.

\bibitem[{\citenamefont{Stecker et~al.}(2016)\citenamefont{Stecker, Scully, and
  Malkan}}]{Stecker:2016fsg}
\bibinfo{author}{\bibfnamefont{F.~W.} \bibnamefont{Stecker}},
  \bibinfo{author}{\bibfnamefont{S.~T.} \bibnamefont{Scully}},
  \bibnamefont{and} \bibinfo{author}{\bibfnamefont{M.~A.}
  \bibnamefont{Malkan}}, \bibinfo{journal}{Astrophys. J.}
  \textbf{\bibinfo{volume}{827}}, \bibinfo{pages}{6} (\bibinfo{year}{2016}),
  \bibinfo{note}{[Erratum: Astrophys. J.863,no.1,112(2018)]},
  \eprint{1605.01382}.

\bibitem[{\citenamefont{Hauser et~al.}(1998)}]{Hauser:1998ri}
\bibinfo{author}{\bibfnamefont{M.~G.} \bibnamefont{Hauser}}
  \bibnamefont{et~al.}, \bibinfo{journal}{Astrophys. J.}
  \textbf{\bibinfo{volume}{508}}, \bibinfo{pages}{25} (\bibinfo{year}{1998}),
  \eprint{astro-ph/9806167}.

\bibitem[{\citenamefont{Matsumoto et~al.}(2005)\citenamefont{Matsumoto,
  Matsuura, Murakami, Tanaka, Freund, Lim, Cohen, Kawada, and
  Noda}}]{Matsumoto:2004dx}
\bibinfo{author}{\bibfnamefont{T.}~\bibnamefont{Matsumoto}},
  \bibinfo{author}{\bibfnamefont{S.}~\bibnamefont{Matsuura}},
  \bibinfo{author}{\bibfnamefont{H.}~\bibnamefont{Murakami}},
  \bibinfo{author}{\bibfnamefont{M.}~\bibnamefont{Tanaka}},
  \bibinfo{author}{\bibfnamefont{M.}~\bibnamefont{Freund}},
  \bibinfo{author}{\bibfnamefont{M.}~\bibnamefont{Lim}},
  \bibinfo{author}{\bibfnamefont{M.}~\bibnamefont{Cohen}},
  \bibinfo{author}{\bibfnamefont{M.}~\bibnamefont{Kawada}}, \bibnamefont{and}
  \bibinfo{author}{\bibfnamefont{M.}~\bibnamefont{Noda}},
  \bibinfo{journal}{Astrophys. J.} \textbf{\bibinfo{volume}{626}},
  \bibinfo{pages}{31} (\bibinfo{year}{2005}), \eprint{astro-ph/0411593}.

\bibitem[{\citenamefont{Tsumura et~al.}(2013)\citenamefont{Tsumura, Matsumoto,
  Matsuura, Sakon, and Wada}}]{Tsumura:2013iza}
\bibinfo{author}{\bibfnamefont{K.}~\bibnamefont{Tsumura}},
  \bibinfo{author}{\bibfnamefont{T.}~\bibnamefont{Matsumoto}},
  \bibinfo{author}{\bibfnamefont{S.}~\bibnamefont{Matsuura}},
  \bibinfo{author}{\bibfnamefont{I.}~\bibnamefont{Sakon}}, \bibnamefont{and}
  \bibinfo{author}{\bibfnamefont{T.}~\bibnamefont{Wada}},
  \bibinfo{journal}{Publ. Astron. Soc. Jap.} \textbf{\bibinfo{volume}{65}},
  \bibinfo{pages}{121} (\bibinfo{year}{2013}), \eprint{1307.6740}.

\bibitem[{\citenamefont{Matsumoto et~al.}(2015)\citenamefont{Matsumoto, Kim,
  Pyo, and Tsumura}}]{Matsumoto:2015fma}
\bibinfo{author}{\bibfnamefont{T.}~\bibnamefont{Matsumoto}},
  \bibinfo{author}{\bibfnamefont{M.~G.} \bibnamefont{Kim}},
  \bibinfo{author}{\bibfnamefont{J.}~\bibnamefont{Pyo}}, \bibnamefont{and}
  \bibinfo{author}{\bibfnamefont{K.}~\bibnamefont{Tsumura}},
  \bibinfo{journal}{Astrophys. J.} \textbf{\bibinfo{volume}{807}},
  \bibinfo{pages}{57} (\bibinfo{year}{2015}), \eprint{1501.01359}.

\bibitem[{\citenamefont{Sano et~al.}(2015)\citenamefont{Sano, Kawara, Matsuura,
  Kataza, Arai, and Matsuoka}}]{Sano:2015bsa}
\bibinfo{author}{\bibfnamefont{K.}~\bibnamefont{Sano}},
  \bibinfo{author}{\bibfnamefont{K.}~\bibnamefont{Kawara}},
  \bibinfo{author}{\bibfnamefont{S.}~\bibnamefont{Matsuura}},
  \bibinfo{author}{\bibfnamefont{H.}~\bibnamefont{Kataza}},
  \bibinfo{author}{\bibfnamefont{T.}~\bibnamefont{Arai}}, \bibnamefont{and}
  \bibinfo{author}{\bibfnamefont{Y.}~\bibnamefont{Matsuoka}},
  \bibinfo{journal}{Astrophys. J.} \textbf{\bibinfo{volume}{811}},
  \bibinfo{pages}{77} (\bibinfo{year}{2015}), \eprint{1508.02806}.

\bibitem[{\citenamefont{Matsuura et~al.}(2017)}]{Matsuura:2017lub}
\bibinfo{author}{\bibfnamefont{S.}~\bibnamefont{Matsuura}}
  \bibnamefont{et~al.}, \bibinfo{journal}{Astrophys. J.}
  \textbf{\bibinfo{volume}{839}}, \bibinfo{pages}{7} (\bibinfo{year}{2017}),
  \eprint{1704.07166}.

\bibitem[{\citenamefont{Kelsall et~al.}(1998)}]{Kelsall:1998bq}
\bibinfo{author}{\bibfnamefont{T.}~\bibnamefont{Kelsall}} \bibnamefont{et~al.},
  \bibinfo{journal}{Astrophys. J.} \textbf{\bibinfo{volume}{508}},
  \bibinfo{pages}{44} (\bibinfo{year}{1998}), \eprint{astro-ph/9806250}.

\bibitem[{\citenamefont{Kashlinsky et~al.}(2004)\citenamefont{Kashlinsky,
  Arendt, Gardner, Mather, and Moseley}}]{Kashlinsky:2004xx}
\bibinfo{author}{\bibfnamefont{A.}~\bibnamefont{Kashlinsky}},
  \bibinfo{author}{\bibfnamefont{R.}~\bibnamefont{Arendt}},
  \bibinfo{author}{\bibfnamefont{J.~P.} \bibnamefont{Gardner}},
  \bibinfo{author}{\bibfnamefont{J.~C.} \bibnamefont{Mather}},
  \bibnamefont{and} \bibinfo{author}{\bibfnamefont{S.~H.}
  \bibnamefont{Moseley}}, \bibinfo{journal}{Astrophys. J.}
  \textbf{\bibinfo{volume}{608}}, \bibinfo{pages}{1} (\bibinfo{year}{2004}),
  \eprint{astro-ph/0401401}.

\bibitem[{\citenamefont{Kohri et~al.}(2017)\citenamefont{Kohri, Moroi, and
  Nakayama}}]{Kohri:2017oqn}
\bibinfo{author}{\bibfnamefont{K.}~\bibnamefont{Kohri}},
  \bibinfo{author}{\bibfnamefont{T.}~\bibnamefont{Moroi}}, \bibnamefont{and}
  \bibinfo{author}{\bibfnamefont{K.}~\bibnamefont{Nakayama}},
  \bibinfo{journal}{Phys. Lett.} \textbf{\bibinfo{volume}{B772}},
  \bibinfo{pages}{628} (\bibinfo{year}{2017}), \eprint{1706.04921}.

\bibitem[{\citenamefont{Kalashev et~al.}(2019)\citenamefont{Kalashev, Kusenko,
  and Vitagliano}}]{Kalashev:2018bra}
\bibinfo{author}{\bibfnamefont{O.~E.} \bibnamefont{Kalashev}},
  \bibinfo{author}{\bibfnamefont{A.}~\bibnamefont{Kusenko}}, \bibnamefont{and}
  \bibinfo{author}{\bibfnamefont{E.}~\bibnamefont{Vitagliano}},
  \bibinfo{journal}{Phys. Rev.} \textbf{\bibinfo{volume}{D99}},
  \bibinfo{pages}{023002} (\bibinfo{year}{2019}), \eprint{1808.05613}.

\bibitem[{\citenamefont{Korochkin et~al.}(2019)\citenamefont{Korochkin,
  Neronov, and Semikoz}}]{Korochkin:2019qpe}
\bibinfo{author}{\bibfnamefont{A.}~\bibnamefont{Korochkin}},
  \bibinfo{author}{\bibfnamefont{A.}~\bibnamefont{Neronov}}, \bibnamefont{and}
  \bibinfo{author}{\bibfnamefont{D.}~\bibnamefont{Semikoz}}
  (\bibinfo{year}{2019}), \eprint{1911.13291}.

\bibitem[{\citenamefont{Acciari et~al.}(2019)}]{Acciari:2019zgl}
\bibinfo{author}{\bibfnamefont{V.~A.} \bibnamefont{Acciari}}
  \bibnamefont{et~al.} (\bibinfo{collaboration}{MAGIC}), \bibinfo{journal}{Mon.
  Not. Roy. Astron. Soc.} \textbf{\bibinfo{volume}{486}}, \bibinfo{pages}{4233}
  (\bibinfo{year}{2019}), \eprint{1904.00134}.

\bibitem[{\citenamefont{Abeysekara et~al.}(2019)}]{Abeysekara:2019ybp}
\bibinfo{author}{\bibfnamefont{A.~U.} \bibnamefont{Abeysekara}}
  \bibnamefont{et~al.} (\bibinfo{collaboration}{VERITAS})
  (\bibinfo{year}{2019}), \eprint{1910.00451}.

\bibitem[{\citenamefont{Levenson et~al.}(2007)\citenamefont{Levenson, Wright,
  and Johnson}}]{Levenson:2007is}
\bibinfo{author}{\bibfnamefont{L.~R.} \bibnamefont{Levenson}},
  \bibinfo{author}{\bibfnamefont{E.~L.} \bibnamefont{Wright}},
  \bibnamefont{and} \bibinfo{author}{\bibfnamefont{B.~D.}
  \bibnamefont{Johnson}}, \bibinfo{journal}{Astrophys. J.}
  \textbf{\bibinfo{volume}{666}}, \bibinfo{pages}{34} (\bibinfo{year}{2007}),
  \eprint{0704.1498}.

\bibitem[{\citenamefont{Matsuoka et~al.}(2011)\citenamefont{Matsuoka, Ienaka,
  Kawara, and Oyabu}}]{Matsuoka:2011hb}
\bibinfo{author}{\bibfnamefont{Y.}~\bibnamefont{Matsuoka}},
  \bibinfo{author}{\bibfnamefont{N.}~\bibnamefont{Ienaka}},
  \bibinfo{author}{\bibfnamefont{K.}~\bibnamefont{Kawara}}, \bibnamefont{and}
  \bibinfo{author}{\bibfnamefont{S.}~\bibnamefont{Oyabu}},
  \bibinfo{journal}{Astrophys. J.} \textbf{\bibinfo{volume}{736}},
  \bibinfo{pages}{119} (\bibinfo{year}{2011}), \eprint{1106.4413}.

\bibitem[{\citenamefont{Pesce et~al.}(1995)\citenamefont{Pesce, Falomo, and
  Treves}}]{pesce1995environmental}
\bibinfo{author}{\bibfnamefont{J.~E.} \bibnamefont{Pesce}},
  \bibinfo{author}{\bibfnamefont{R.}~\bibnamefont{Falomo}}, \bibnamefont{and}
  \bibinfo{author}{\bibfnamefont{A.}~\bibnamefont{Treves}},
  \bibinfo{journal}{The Astronomical Journal} \textbf{\bibinfo{volume}{110}},
  \bibinfo{pages}{1554} (\bibinfo{year}{1995}).

\bibitem[{\citenamefont{Falomo et~al.}(2014)\citenamefont{Falomo, Pian, and
  Treves}}]{Falomo:2014yya}
\bibinfo{author}{\bibfnamefont{R.}~\bibnamefont{Falomo}},
  \bibinfo{author}{\bibfnamefont{E.}~\bibnamefont{Pian}}, \bibnamefont{and}
  \bibinfo{author}{\bibfnamefont{A.}~\bibnamefont{Treves}},
  \bibinfo{journal}{Astron. Astrophys. Rev.} \textbf{\bibinfo{volume}{22}},
  \bibinfo{pages}{73} (\bibinfo{year}{2014}), \eprint{1407.7615}.

\bibitem[{\citenamefont{Falomo et~al.}(1993)\citenamefont{Falomo, Pesce, and
  Treves}}]{falomo1993environment}
\bibinfo{author}{\bibfnamefont{R.}~\bibnamefont{Falomo}},
  \bibinfo{author}{\bibfnamefont{J.~E.} \bibnamefont{Pesce}}, \bibnamefont{and}
  \bibinfo{author}{\bibfnamefont{A.}~\bibnamefont{Treves}},
  \bibinfo{journal}{The Astrophysical Journal} \textbf{\bibinfo{volume}{411}},
  \bibinfo{pages}{L63} (\bibinfo{year}{1993}).

\bibitem[{\citenamefont{Farina et~al.}(2016)\citenamefont{Farina, Fumagalli,
  Decarli, and Fanidakis}}]{Farina:2015rwa}
\bibinfo{author}{\bibfnamefont{E.~P.} \bibnamefont{Farina}},
  \bibinfo{author}{\bibfnamefont{M.}~\bibnamefont{Fumagalli}},
  \bibinfo{author}{\bibfnamefont{R.}~\bibnamefont{Decarli}}, \bibnamefont{and}
  \bibinfo{author}{\bibfnamefont{N.}~\bibnamefont{Fanidakis}},
  \bibinfo{journal}{Mon. Not. Roy. Astron. Soc.}
  \textbf{\bibinfo{volume}{455}}, \bibinfo{pages}{618} (\bibinfo{year}{2016}),
  \eprint{1510.01779}.

\bibitem[{\citenamefont{Carilli and Taylor}(2002)}]{Carilli:2001hj}
\bibinfo{author}{\bibfnamefont{C.~L.} \bibnamefont{Carilli}} \bibnamefont{and}
  \bibinfo{author}{\bibfnamefont{G.~B.} \bibnamefont{Taylor}},
  \bibinfo{journal}{Ann. Rev. Astron. Astrophys.}
  \textbf{\bibinfo{volume}{40}}, \bibinfo{pages}{319} (\bibinfo{year}{2002}),
  \eprint{astro-ph/0110655}.

\bibitem[{\citenamefont{Meyer et~al.}(2014)\citenamefont{Meyer, Montanino, and
  Conrad}}]{Meyer:2014epa}
\bibinfo{author}{\bibfnamefont{M.}~\bibnamefont{Meyer}},
  \bibinfo{author}{\bibfnamefont{D.}~\bibnamefont{Montanino}},
  \bibnamefont{and} \bibinfo{author}{\bibfnamefont{J.}~\bibnamefont{Conrad}},
  \bibinfo{journal}{JCAP} \textbf{\bibinfo{volume}{1409}}, \bibinfo{pages}{003}
  (\bibinfo{year}{2014}), \eprint{1406.5972}.

\bibitem[{\citenamefont{Pshirkov et~al.}(2016)\citenamefont{Pshirkov, Tinyakov,
  and Urban}}]{Pshirkov:2015tua}
\bibinfo{author}{\bibfnamefont{M.~S.} \bibnamefont{Pshirkov}},
  \bibinfo{author}{\bibfnamefont{P.~G.} \bibnamefont{Tinyakov}},
  \bibnamefont{and} \bibinfo{author}{\bibfnamefont{F.~R.} \bibnamefont{Urban}},
  \bibinfo{journal}{Phys. Rev. Lett.} \textbf{\bibinfo{volume}{116}},
  \bibinfo{pages}{191302} (\bibinfo{year}{2016}), \eprint{1504.06546}.

\bibitem[{\citenamefont{Jansson and Farrar}(2012)}]{Jansson:2012rt}
\bibinfo{author}{\bibfnamefont{R.}~\bibnamefont{Jansson}} \bibnamefont{and}
  \bibinfo{author}{\bibfnamefont{G.~R.} \bibnamefont{Farrar}},
  \bibinfo{journal}{Astrophys. J.} \textbf{\bibinfo{volume}{761}},
  \bibinfo{pages}{L11} (\bibinfo{year}{2012}), \eprint{1210.7820}.

\bibitem[{\citenamefont{Meyer}(2013)}]{meyer2013opacity}
\bibinfo{author}{\bibfnamefont{M.}~\bibnamefont{Meyer}}, \bibinfo{type}{Tech.
  Rep.} (\bibinfo{year}{2013}).

\bibitem[{\citenamefont{Wilks}(1938)}]{wilks1938large}
\bibinfo{author}{\bibfnamefont{S.~S.} \bibnamefont{Wilks}},
  \bibinfo{journal}{The Annals of Mathematical Statistics}
  \textbf{\bibinfo{volume}{9}}, \bibinfo{pages}{60} (\bibinfo{year}{1938}).

\bibitem[{\citenamefont{Acharya et~al.}(2013)}]{Acharya:2013sxa}
\bibinfo{author}{\bibfnamefont{B.~S.} \bibnamefont{Acharya}}
  \bibnamefont{et~al.} (\bibinfo{collaboration}{CTA Consortium}),
  \bibinfo{journal}{Astropart. Phys.} \textbf{\bibinfo{volume}{43}},
  \bibinfo{pages}{3} (\bibinfo{year}{2013}).

\bibitem[{\citenamefont{Cao}(2010)}]{Cao:2010zz}
\bibinfo{author}{\bibfnamefont{Z.}~\bibnamefont{Cao}}
  (\bibinfo{collaboration}{LHAASO}), \bibinfo{journal}{Chin. Phys.}
  \textbf{\bibinfo{volume}{C34}}, \bibinfo{pages}{249} (\bibinfo{year}{2010}).

\end{thebibliography}

\begin{appendix}
\end{appendix}

\end{document}